\documentclass{article}

\usepackage{arxiv}
\usepackage[export]{adjustbox}
\usepackage{amssymb}
\usepackage{url}
\usepackage{cite}
\usepackage[utf8]{inputenc}
\usepackage{amsmath}
\usepackage{hyperref}
\usepackage{enumitem}
\usepackage{graphicx}
\usepackage{siunitx}
\usepackage{array}
\usepackage{mathtools}
\usepackage{tcolorbox}
\usepackage{tikz}
\usetikzlibrary{decorations.pathreplacing}
\usepackage{pifont}
\usepackage{caption}
\usepackage{subcaption}
\usepackage{varwidth}
\usepackage{xcolor,colortbl}
\usepackage{listings}

\lstdefinestyle{mystyle}{
    backgroundcolor=\color{lightgray!40},   
    commentstyle=\color{green},
    keywordstyle=\color{blue},
    numberstyle=\tiny\color{gray},
    stringstyle=\color{red},
    basicstyle=\ttfamily\footnotesize,
    breakatwhitespace=false,         
    breaklines=true,                 
    captionpos=b,                    
    keepspaces=true,                 
    numbers=left,                    
    numbersep=5pt,                  
    showspaces=false,                
    showstringspaces=false,
    showtabs=false,                  
    tabsize=2
}

\begin{document}
\pagenumbering{gobble}


\title{Security Assessment of Mobile Banking Apps in West African Economic and Monetary Union}

\author{
Alioune Diallo \\
\textit{SnT/TruX} \\
  \textit{University of Luxembourg} \\
  alioune.diallo@uni.lu \\
   \And
    Aicha WAR\\
    \textit{SnT/TruX} \\
    \textit{University of Luxemboug} \\
    Kirchberg, Luxembourg \\
    aicha.war@uni.lu \\
    \And
    Moustapha Awwalou DIOUF\\
   \textit{SnT/TruX} \\
    \textit{University of Luxemboug} \\
    Kirchberg, Luxembourg \\
    moustapha.diouf@uni.lu\\
    \And
    Jordan Samhi \\
  \textit{Software Research Group}\\
  \textit{CISPA Helmholtz Center for Information Security} \\
  Saarbrücken, Germany \\
    jordan.samhi@cispa.de \\
  \And
    Steven ARZT\\
    \textit{Secure Software Engineering} \\
    \textit{Fraunhofer SIT} \\
    Darmstadt, Germany \\
     steven.arzt@sit.fraunhofer.de \\
     \And
     Tegawendé F. BISSYANDE\\
        \textit{SnT/TruX} \\
        \textit{University of Luxemboug} \\
        Kirchberg, Luxembourg \\
        tegawende.bissyande@uni.lu
      \And
        Jacque KLEIN\\
        \textit{SnT/TruX} \\
        \textit{University of Luxemboug} \\
        Kirchberg, Luxembourg \\
        jacques.klein@uni.lu
}

\maketitle

\begin{abstract}
Mobile banking adoption is soaring in Africa, particularly within the West African Economic and Monetary Union (WAEMU) states.
These countries, characterized by widespread smartphone usage, have witnessed banks and financial institutions introducing mobile banking applications. 
These apps empower users to perform transactions such as money transfers, bill payments, and account inquiries anytime, anywhere.
However, this proliferation of mobile banking apps also raises significant security concerns. 
Poorly implemented security measures during app development can expose users and financial institutions to substantial financial risks through increased vulnerability to cyberattacks. 
Our study evaluated fifty-nine WAEMU mobile banking apps using static analysis techniques. 
These mobile banking apps were collected from the 160 banks and financial institutions of the eight WAEMU countries listed on the Central Bank of West African States (BCEAO) website.
We identified security-related code issues that could be exploited by malicious actors.
We investigated the issues found in the older versions to track their evolution across updates. 
Additionally, we identified some banks from regions such as Europe, the United States, and other developing countries and analyzed their mobile apps for a security comparison with WAEMU banking apps.
Key findings include: (1) WAEMU apps exhibit security issues introduced during development, posing significant risks of exploitation; (2) Despite frequent updates, underlying security issues often persist; (3) Compared to banking apps from developed and developing countries, WAEMU apps exhibit fewer critical security issues; and (4) Apps from banks that are branches of other non-WAEMU banks often inherit security concerns from their parent apps while also introducing additional issues unique to their context.
Our research underscores the need for robust security practices in WAEMU mobile banking app development to enhance user safety and trust in financial services.
\\
\end{abstract}


\section{Introduction}
 The financial sector plays a pivotal role in the economic development of any country, with digital financial services, including mobile money and mobile banking, emerging as powerful catalysts for this progress, particularly in Sub-Saharan African nations. 
The widespread adoption of mobile banking is a global phenomenon, with Sub-Saharan Africa leading the way in its innovative implementation~\cite{bei}. 
This surge can be attributed mainly to the remarkable proliferation of smartphones, a trend projected to continue its upward trajectory. 
Nowadays, approximately 60\% of people have smartphones worldwide~\cite{ashTurner}.
By 2022, smartphone adoption in Sub-Saharan Africa had reached an impressive 51\%, with projections indicating a substantial rise to 87\% by 2030~\cite{osiris}.

Recognizing the strategic advantage of widespread smartphone penetration, banks and financial institutions within the West African Economic and Monetary Union (WAEMU) region are leveraging this technology to extend banking services. 
This leap serves a dual purpose: broadening access to banking services for millions and empowering existing account holders with remote control over their finances, facilitating transactions and balance checks. 
In WAEMU, nearly all banks and financial institutions have embraced mobile applications, reaching millions reliant on these platforms for daily financial transactions.

However, the ubiquity of mobile banking applications is accompanied by recurring security concerns, posing challenges for users and institutions alike. 
Reports highlight, for example, the persistence of sensitive keys hard-coded into financial mobile applications across Africa~\cite{appRoov}, underscoring the potential for significant financial losses. 
Despite the growing reliance on these applications in developing countries, including in the WAEMU, comprehensive studies on their security remain scarce.
A literature review reveals limited studies on mobile financial applications in developing nations, with no research specific to the WAEMU region~\cite{diallo2024security}.


Various approaches have been explored to assess the security issues of mobile banking apps, ranging from investigating their origins~\cite{Ansong} to conducting forensic examinations and vulnerability assessments~\cite{Uduimoh, Osho}. 
However, there remains a gap in research addressing these concerns comprehensively.

Against this backdrop, our study extensively investigates the security landscape of mobile banking applications in WAEMU countries. 
Our research aims to identify prevalent security issues, establish a comprehensive threat model, analyze the evolution of security across different app versions, and compare the security posture of WAEMU banking apps with those from the European Union (EU), the United States (US), and other developing countries (ODC). 
Moreover, some WAEMU financial institutions are branches (children) of other institutions (parents). 
We explore the security of their mobile banking apps and determine if they inherit parent app security issues.

The contributions of this study are as follows:
\begin{itemize}
    \item Highlighting the most common security problems affecting WAEMU banking apps.
    \item Providing possible ways to exploit app vulnerabilities to understand the risks better. 
    \item Offering an assessment of the security evolution in these apps, including whether concerns are effectively addressed in subsequent updates.
    \item Offering a comparative analysis of the security posture of WAEMU banking apps in the global context, comparing them with the applications used by leading African and International banks.
    \item Shedding light on the security problems between branch and parent apps.
\end{itemize}

These findings underscore the imperative of enhancing mobile banking app security to safeguard users and institutions in an increasingly digital financial landscape.

We present the rest of the study as follows. 
Section~\ref{motivation} presents the motivation of the study.
Section~\ref{backgrnd} presents the background by explaining the concepts of mobile banking and security code smells.
Section~\ref{method} describes the detailed methodology followed in this paper.
We report the results in Section~\ref{result} and discuss them in Section~\ref{discuss}.
We discuss the threat to validity in Section~\ref{threat} and the related works in Section~\ref{related}.
Finally, we conclude the SLR in Section~\ref{concl}.
\\
\\
\noindent
\textbf{Artifacts.}
We release all of our artifacts (e.g., list of mobile apps, detailed results, etc.):
\begin{center}
\url{https://github.com/liounea/Data_for_WAEMU_Apps_Papers}
\end{center}
\section{Motivation of this study}
\label{motivation}

Mobile banking is a reality across the world.
With the wide use of Smartphones in Sub-Saharan Africa~\cite{osiris}, banks and financial institutions leverage this technology to enhance access to banking services by developing mobile banking applications (MBAs).
Through mobile banking, financial services, and people's money have become at risk with the increased attack surfaces.
In Africa, such as WAEMU, the use of mobile banking is increasing steadily.
Nearly all banks and financial institutions from WAEMU countries have deployed MBAs, reaching millions of users who rely on these applications for their daily financial activities.

However, there are security challenges related to MBAs. 
Indeed, the low literacy rate in Sub-Saharan Africa~\cite{rateLiteracy} poses a significant challenge in terms of whether these mobile banking apps are used. 
For example, the lack of literacy can lead users to be victims of fraud or phishing attacks when using apps. 
Since MBAs process sensitive data, they are not without risks and vulnerabilities. 
These can pose recurrent security problems. 
For instance, sensitive keys persist within mobile financial applications from Africa~\cite{appRoov}. 
Such security issues can result in significant financial losses for both users and financial institutions.





These known risks motivate our study.
It will contribute to raising awareness and securing MBAs in WAEMU countries.
As mentioned in a prior work~\cite{10.1145/3001913.3001919}: "Where there is money, there must be security," users must be reassured, and banks should care about users' security through MBAs to encourage their adoption and facilitate financial inclusion.

\section{Background}
\label{backgrnd}
\subsection{What is mobile banking?}
In this study, we excluded mobile money apps that are not necessarily related to a bank account.
We indeed focus on mobile banking which offers services that allow customers to use their bank account through mobile equipment~\cite{pousttchi2004assessment}.
These services could be bank transfers from one account to another, balance checking, or bill payments.
In general, banks develop applications that could be used on Smartphones to facilitate the use of their services.
Those applications are called mobile banking applications or mobile banking apps (shortly MBAs here).

\subsection{What are security code smells and vulnerability?}
Security code smells (code smells or smell) are security issues coming from a bad implementation and poor design in software~\cite{tufano2017and, elkhail2019relating}.
They are not necessarily exploitable. 
However, they could have consequences and could compromise the security and privacy of software users~\cite{ghafari2017security}. 
In this case, they are called vulnerabilities.

\section{Methodology}
\label{method}
This work assesses the security of MBAs from financial institutions in the WAEMU countries.
This section describes the Research Questions (RQs) and the app collection and analysis process.

\subsection{Research questions}

Our study aimed to provide a comprehensive understanding of the security landscape surrounding mobile banking apps in the WAEMU region. To achieve this, we formulated 4 RQs.

\textbf{RQ1: To what extent do mobile banking apps from WAEMU present critical security issues?}

We have scanned previously collected banking apps for vulnerabilities to address this RQ.
Several critical issues have been found in WAEMU banking apps.
We have confirmed some of these critical issues as vulnerabilities by manual verification and have examined real-world attack scenarios for some of these critical vulnerabilities.
This research question reveals that banking apps in the WAEMU are highly vulnerable and can be easily compromised.
Hence, practitioners must recognize these issues and address them to make mobile banking safer in the WAEMU.



\textbf{RQ2: How do security issues evolve in WAEMU bank apps?}
Given the ever-evolving landscape of technology, the discovery of new security issues and vulnerabilities in mobile apps is a persistent concern. 
This question investigates the responsiveness of WAEMU banks to these security concerns. 
It investigates the practice of proposing and implementing regular updates to address these concerns and scrutinizes whether these updates effectively rectify issues in subsequent app versions.

\textbf{RQ3: How vulnerable are WAEMU banking apps compared to apps from 1) the top 20 EU banks, 2) the top 20 US banks, and 3) the top 20 other developing countries’ banks?}
Our study extends its purview to include a comparative analysis of WAEMU banking apps against those of top European Union and United States banks and top banking apps from other (i.e., non-WAEMU) developing countries. 
This comparative assessment serves several critical purposes: it elucidates whether these apps share similar security issue profiles, discerns the disparities in vulnerability severity and distribution, and identifies factors that render WAEMU banking apps more or less vulnerable than their international counterparts.

\textbf{RQ4: How vulnerable are child banking apps compared to parent banking apps?}
Some WAEMU banks are branches of international banks. 
We carry out this study to investigate if the apps of those banks (child apps) inherit the problems of the respective African and international parent banks.

Through the answers to these research questions, we provide a comprehensive examination of the security status of mobile banking apps in the WAEMU region, offering actionable insights and potential avenues for enhancing the security of these applications to safeguard the interests of both users and financial institutions.

\subsection{App selection process}

\subsubsection{Collecting WAEMU banking apps}
We meticulously considered all the 160 banks and financial institutions across the eight WAEMU countries\footnote{The eight countries are: Benin, Burkina Faso, Côte D'Ivoire, Guinea-Bissau, Mali, Niger, Senegal, and Togo} listed on the Central Bank of West African States (BCEAO) website~\cite{Bceao}.
Subsequently, we gathered a comprehensive dataset comprising only fifty-nine (59) Android banking applications from Google Play.
We do not have the same number of applications as institutions (160) because several institutions have the same mobile app, and others have none.  
The oldest available APKs have been collected from December 15 to December 16, 2022.
Our decision to focus exclusively on the Android platform stems from its predominant usage in Africa.
As indicated by statistics~\cite{GS}, the Android operating system lead the market share in Africa in November 2022, with a rate of 83.87\% .

To better understand the security implications of these findings, we manually analyzed the apps' source code.
This manual examination facilitated the establishment of potential real-world attack vectors for exploiting the identified vulnerabilities.

\subsubsection{Selection of old versions of WAEMU apps}
We meticulously collected historical versions of these apps to investigate the evolution of security issues across different versions of WAEMU banking apps. 
As previously mentioned, our study encompasses 59 WAEMU apps, comprising the newest available APKs.
These served as reference points for analyzing the evolution of security features.
We relied on AndroZoo~\cite{Allix:2016:ACM:2901739.2903508} to procure the historical APKs for all 59 apps.
Among the 59 apps, some (approximately 20\%) do not have more than 2 versions.
AndroZoo provides extensive metadata for each APK, of which the most important for our study are the SHA256 hash, package name, and the date of addition (the date on which the corresponding APK is added to AndroZoo).
We differentiate two versions of the same app if they have the same package name and different SHA256 hashes.
The oldest one among the two versions is determined considering the dates of addition to AndroZoo.
AndroZoo regularly crawls the Google Play Store. Hence, the date of addition to AndroZoo is representative of the age of the app in the store.

To facilitate a comprehensive understanding of security evolution, we aimed to select five versions for each app, including the oldest APK. The versions were chosen at an interval of two months, at least between two consecutive versions. 
Several apps have an important number of versions.
For those apps, we considered an interval of one year at least to choose versions. 
Some apps did not have sufficient historical versions, so we did not need to exclude versions in such cases.

Consequently, our final dataset comprises apps with varying numbers of versions, ranging from 2 to 5, inclusive of the reference version. 
This meticulous approach enables a nuanced analysis of security trends and evolution across different iterations of WAEMU mobile banking applications.

\subsubsection{Selection of European Union's and United States apps}
For the European Union (EU) and United States (US) banking apps, we have selected the top 20 banks for each of them based on their total asset value~\cite{mYuen, gaby}. Each of the chosen banks can have many mobile apps due to their presence in many countries. 
Thus, we identified the package names of these banks' apps from their respective headquarters cities, as the package name serves as a unique identifier for each app.

\subsubsection{Selection of apps from other developing countries}
Apps from developing countries other than WAEMU countries have been selected, too.
The goal is to compare the security of mobile banking apps within similar development contexts.
To respect the same process as for developed countries,  we selected the top 20 banks from developing countries, excluding WAEMU members.
The selection criteria were based on the 2023 bank rankings provided by Brandirectory on its website~\cite{bDir}, which ranks banks globally according to their total assets.
Brandirectory's comprehensive ranking lists the top 500 banks in 2023, from which we identified the 20 most valuable banks in developing countries, excluding those from WAEMU.

\subsection{Automated analysis tool}
\label{vusc}

To analyze our collected MBAs, we employed a robust vulnerability scanner tool called VUSC\footnote{\url{https://www.sit.fraunhofer.de/en/offers/projekte/vusc/}}.
VUSC is a commercial tool built upon the Dexpler~\cite{10.1145/2259051.2259056} and Soot framework~\cite{lam2011soot}, as detailed in prior research~\cite{10.1145/3540250.3549091}. 
This tool translates the app's bytecode into a more analyzable format, such as Jimple code~\cite{vallee1998jimple}, facilitating static analysis to uncover potential vulnerabilities.
Furthermore, VUSC utilizes FlowDroid~\cite{FlowDroid} to conduct data flow analysis, identifying potential data leakage points within the app.
It is essential to note that the results generated by VUSC primarily comprise Security-Related Code Smells (SRCSs), i.e., all of them are not necessarily exploitable, as elucidated in the literature~\cite{10.1145/3540250.3549091}.

\section{Empirical Results}
\label{result}

This section presents the result of our empirical study.

\begin{figure}[t!]
  \centering
  \includegraphics[width=3.4in]{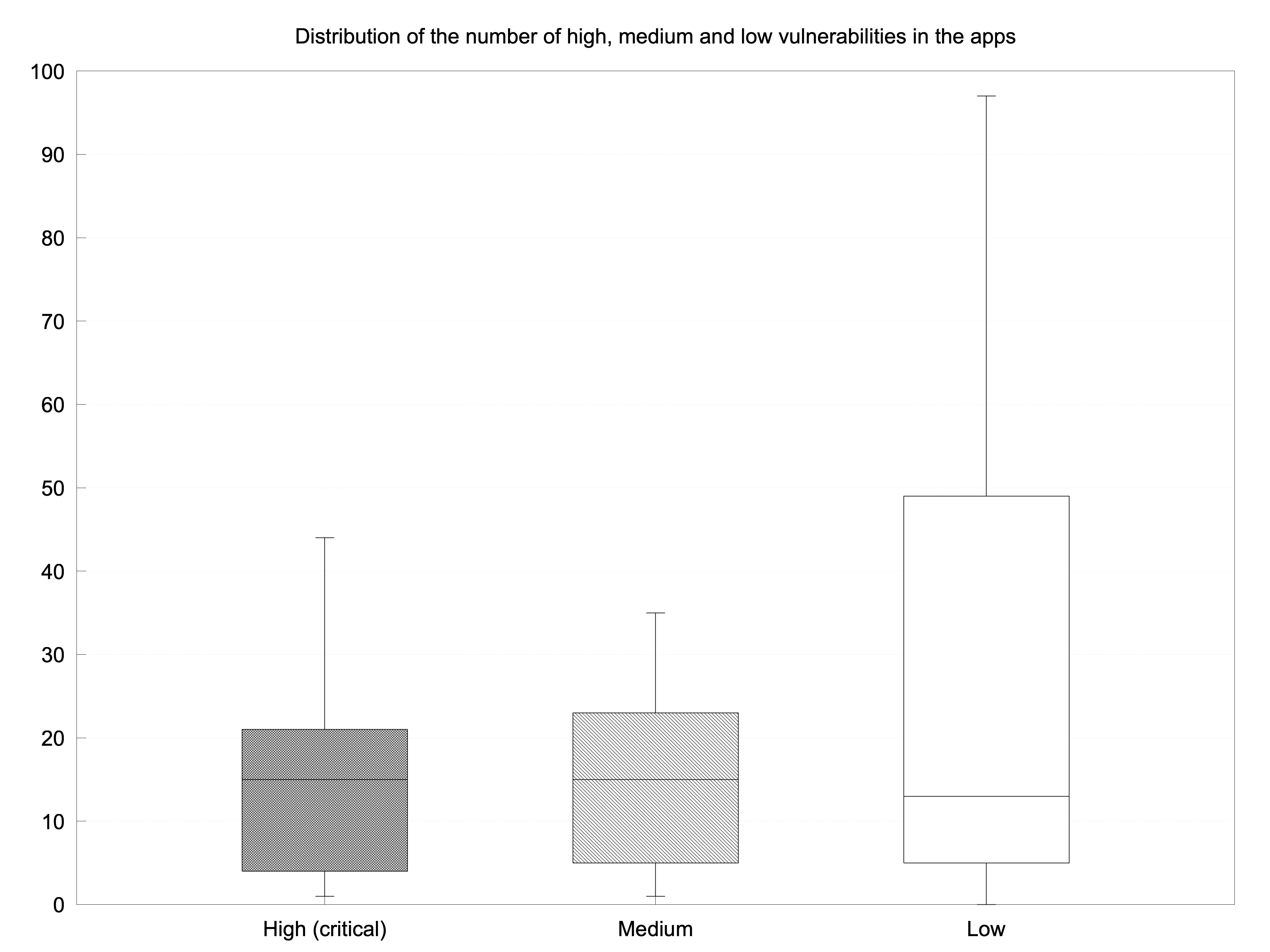}
  \caption{Overview of the number of vulnerabilities found in WAEMU banking apps.}
  \label{fig:hml}
\end{figure}

As presented above, we have collected and successfully analyzed 59 MBAs from the WAEMU countries. 
The VUSC scanner has reported numerous potential security issues for MBAs in WAEMU countries. As these issues are not yet validated, we call them SRCSs, a term coined in previous research~\cite{10.1145/3540250.3549091}. We grouped these SRCSs into three categories: high, medium, and low, as shown in Fig.~\ref{fig:hml}.
\subsection{WAEMU banking app vulnerabilities}

\begin{figure}[t]
  \centering
  \includegraphics[width=3.4in]{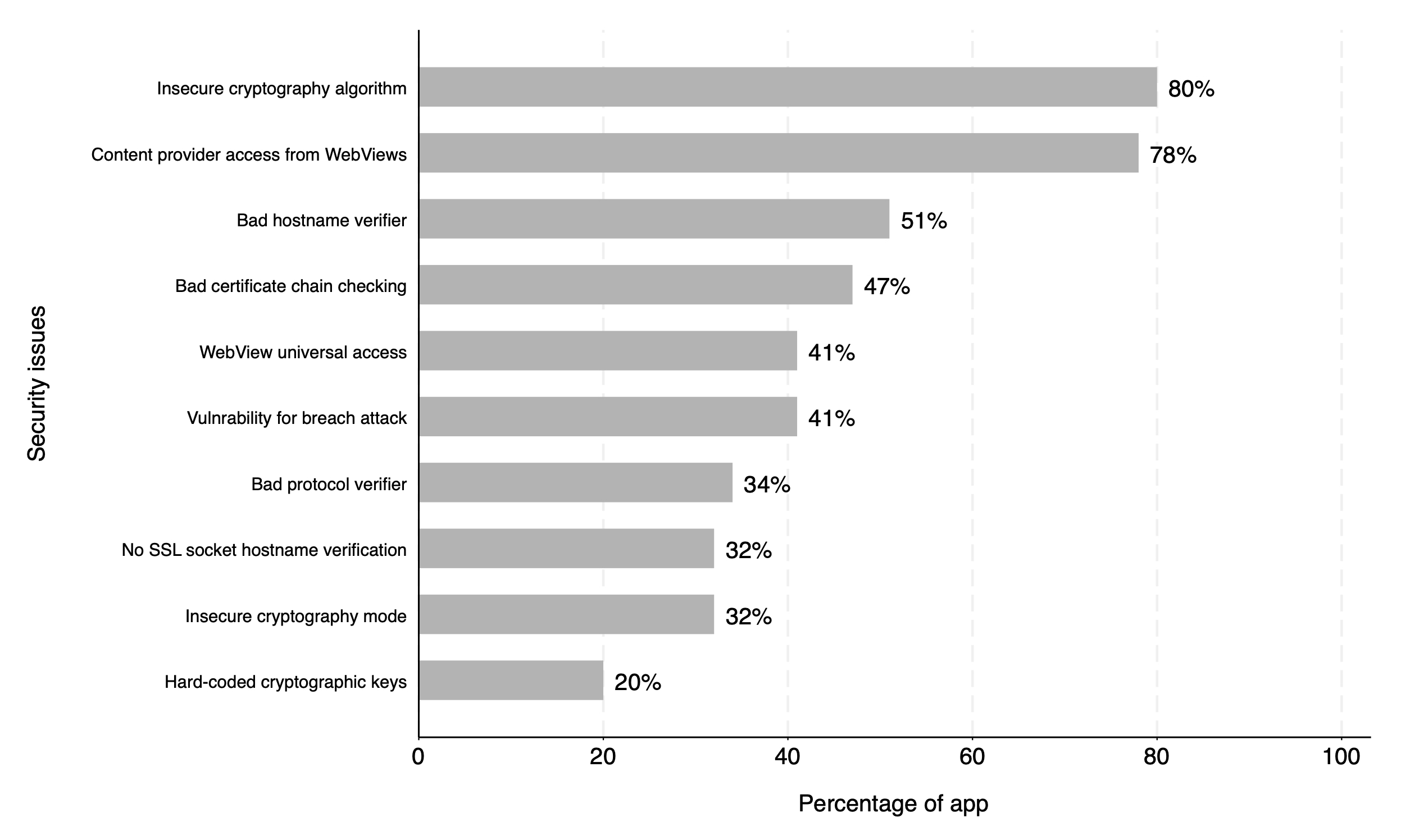}
  \caption{Top ten of the most common critical security issues found in the WAEMU banking apps and the percentage of apps for each vulnerability.}
  \label{fig:top10}
\end{figure}

In the remainder of this work, due to the high number of findings across all apps, we focus on \textbf{the most critical ones, i.e., the ones labeled as high severity by the VUSC scanner.} The results show that each app has at least one possible security issue, and half have over fifteen SRCSs.
As shown in Fig.~\ref{fig:top10}, the ten most common SRCSs in the WAEMU banking apps affect between 20 to 80\% MBAs.
We next describe these highly prevalent SRCSes in detail.

\textbf{Use of insecure cryptographic algorithm} (80\%). 
Cryptographic algorithms may be used for tasks such as encryption/decryption, electronic signatures and the verification thereof, computation of hash functions, and other security-critical functions. Using an insecure or outdated algorithm compromises the respective security property. In the case of encryption, attackers with enough computational power may be able to retrieve the plaintext without knowing the key via cryptanalysis or brute-force attacks.

\textbf{Content provider access from WebViews} (78\%).
Apps may display web content in a component called \texttt{WebView}. Allowing this \texttt{WebView} to interact with content providers inside the same app can be a security risk, especially if the \texttt{WebView} may be tricked into displaying untrusted content or executing untrusted JavaScript code.

\textbf{Bad hostname verifier} (51\%).
When an Android app communicates with a server over HTTPS (SSL/TLS), it typically relies on X.509 certificates to authenticate the server and establish a secure connection. During the SSL/TLS handshake, an app must verify that the server's hostname matches the hostname(s) listed in the X.509 certificate the server presents. By default, this check is handled by the operating system. Apps may, however, implement their own verifier, e.g., to accept debug certificates during development or to cater to special use cases. However, if bad practices or implementation mistakes remain in the productive versions, adversaries may be able to perform man-in-the-middle (MiTM) attacks. If successful, such an attack allows intercepting and modifying all network data between the app and the bank's server.

\textbf{Bad certificate chain checking} (47\%).
Like hostname verification, this vulnerability also focuses on improper handling of the TLS handshake. Proper certificate chain checking involves verifying that the server's certificate is issued by a trusted CA, checking for certificate revocation status, and verifying the entire certificate chain up to a trusted root CA.
However, if the app fails to check the certificate chain properly, it could potentially trust a certificate that has not been issued by a trusted certificate authority (CA) or that has been tampered with. As explained above, this could compromise the user’s security via a man-in-the-middle attack.

\textbf{WebView universal access} (41\%).
The app configures a WebView such that JavaScript code loaded from files can access any other resource (regardless of its type), effectively completely disabling the same origin policy for code executed from a JS file.
This configuration allows JavaScript code to access content providers and similar Android resources.
Consequently, this vulnerability can lead to information disclosure from arbitrary resources.

\textbf{Vulnrability for breach attack} (41\%).
This is related to HTTP compression.
If an app enables compression for an HTTP(s) connection when transferring sensitive data, an attacker can get information about the corresponding response by simply injecting plaintext into the request.

\textbf{Bad protocol verifier} (34\%).
SSL/TLS protocol protects the app's data by encrypting the connection between the client and server.
If protocols considered insecure, such as SSL(v1), SSLv2, SSLv3, TLSv1, and TLSv1.1, are used, this can make connections vulnerable to exploits.

\textbf{No SSL socket hostname verification} (32\%).
SSL socket factory with hostname verification allows connections to be established by ensuring that the certificate presented by the server matches the expected hostname.
If this is not done, it can lead to a vulnerability allowing MiTM attacks.

\textbf{Insecure cryptography mode} (32\%).
A cipher mode applies a cryptographic algorithm to encrypt or decrypt data larger than a single block.
An insecure cipher mode such as ECB may enable attackers to decrypt data based on patterns that remain visible in the ciphertext. In the context of banking apps, this may lead to serious data leaks.

\textbf{Hard-coded cryptographic key} (20\%). 
A cryptographic key (\textbf{key}) is a secret that is used to encrypt or decrypt data (when using symmetric ciphers) or to generate signatures. If the attacker knows the key, they can decrypt the data or forge signatures. Keys should never be hard-coded into apps because attackers can decompile the app to extract the key. Since all users have the same app with the same key, the attacker may target all users and decrypt their data.

\begin{figure}[t!]
    \centering
    \includegraphics[width=3.4in]{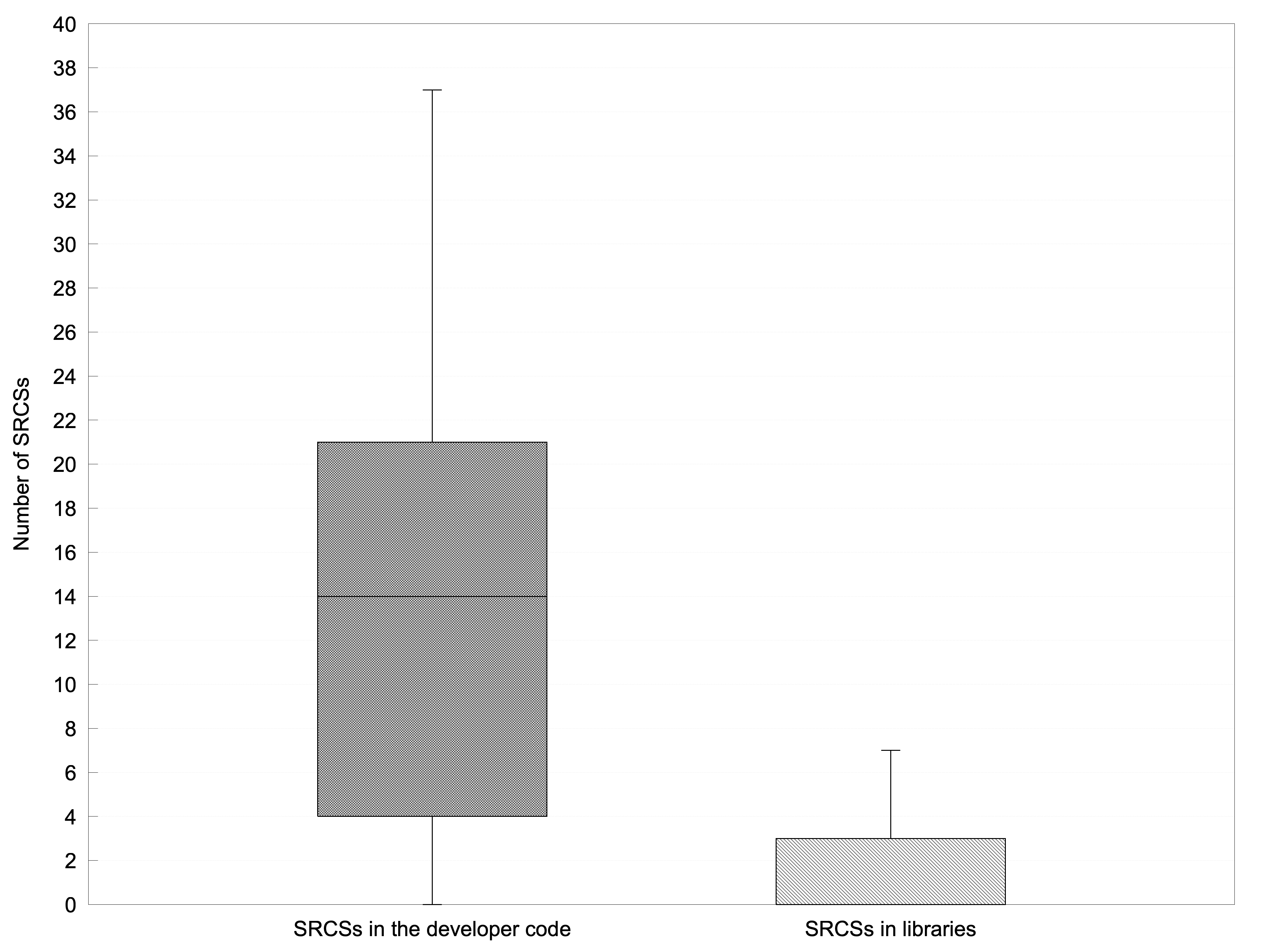}
    \caption{SRCSs in developer code and libraries.}
    \label{fig:devLib}
\end{figure}

Furthermore, we have identified the components with the most critical SRCSs inside the app code to understand if these SRCSs stem from the app developer's code or libraries used by the apps with the help of AndroLibZoo~\cite{10.1145/3643991.3644866}.
As presented in Fig.~\ref{fig:devLib}, half of the apps have no SRCSs from their used libraries, i.e., all SRCSs stem from the original app code. Further, 75\% of the apps have at least four SRCSs in the developer's code. The most critical SRCSes have been found in the developer's code.

To validate the results yielded by the scanner, we manually analyzed a representative sample of SRCSs in our set of MBAs.
We confirm that 10 of the 21 SRCSs manually checked can be exploited; thus, they are vulnerabilities.
Among those 10, 6 are among the top ten most common security issues.
Table~\ref{tab:threat_model} shows these vulnerabilities and possible ways to exploit them.
\begin{table*}[t!]
  \caption{Description of vulnerabilities and possible exploit cases.}
  \label{tab:threat_model}
  \begin{adjustbox}{width=.85\linewidth,center}
  \begin{tabular}{l||ll}
    \textbf{Security issues}&\textbf{Description}&\textbf{Possible way to exploit}\\
    \hline
    &A cryptographic algorithm is used to encrypt and decrypt&\\
    Use of insecure&data  by defining rules. Apps use insecure algorithms, such&\\
    cryptographic&as SHA1, MD5, etc., in encrypting and decrypting sensitive&By using dictionary attacks, an attacker can\\
    algorithm&data. Insecure use of cryptography may enable attackers&easily crack the hash using these hash\\
    (CWE-327, CWE-328)&to access sensitive information which is supposed to be&functions.\\
    &protected. Using insecure algorithms can lead to reputation&\\
    &damage, data loss, etc.&\\
    \hline
    &SSL/TLS HostnameVerify refers to the process of validating & By using MiTM, attackers can spoof the SSL\\
    &that the hostname of the server matches the hostname(s) listed&server via a valid certificate for a different\\
    Use of a bad&in the X.509 certificate presented by the server during the &host,  leading to potential security breaches\\
    hostname verify&SSL/TLS handshake. The SSL/TLS HostnameVerify process is&\\
     (CWE-295)&not implemented correctly in some apps, allowing&\\
     &certificates with incorrect or mismatched hostnames to be&\\
   &accepted. This can lead to data breaches.&\\
    \hline
   &Certificate chain checking ensures the validity of SSL certificates&An attacker needs to be in a MiTM position to\\
    Bad certificate chain&presented by the server when communicating with the app. The&intercept the communication. When the victim\\
    checking (CWE-297)&certificate must come from a trusted certificate authority (CA).&uses the vulnerable app, the attacker can intercept\\
    &Apps fail to check the certificate chain properly.&and modify the app and server communication.\\
    \hline
    No SSL socket&Apps create sockets to connect to specified hosts at specified ports.&Attacker can perform a MiTM attack to intercept\\
    hostname verification&However, many do not perform a hostname verification when&communications and decrypt and modify data\\
     (CWE-297)&establishing connections.&transferred for malicious purposes.\\
    \hline
    &Apps use insecure cryptography cipher mode, such as ECB.&Since the same inputs result in the same ciphertext,\\
    Insecure cryptography&Using insecure cipher modes may enable attackers to decrypt&an attacker can use a chosen  plaintext attack (CPA)\\
    mode (CWE-327)&or tamper data, which can lead to serious security breaches&to guess the encrypted data.\\
    &and data leaks. With ECB, the same inputs result in the&\\
    &same ciphertext.&\\
    \hline
    & A cryptographic key (\textbf{key}) is used to generate a secret key&By reverse-engineering the app, an attacker\\
    Hard-coded&from a SecretKeySpec method to encrypt and decrypt data.&can get the hard-coded key and use it to\\
    cryptographic&If “\textbf{key}” is known, attackers can deduce the secret key to&decrypt any sensitive data that is stored or\\
    key (CWE-321)&decrypt data. Developers often hard-code this key into the&transmitted using the application or device.\\
    &app code.&\\
    \hline
   &The initialization vector is used to randomize cipher&since the same inputs result in the same\\
   Use of static&texts with the same input. When a static IV is used,&ciphertext when using static IV, an\\
   initialization vector&the same inputs result in the same ciphertext, at&attacker can use a chosen plaintext attack\\
   (CWE-1204)&least in parts. As such, a new IV should be selected&(CPA) to guess the encrypted data. \\
   &randomly before each use. The IV is considered public&\\
   &information.&\\
    \hline
    &To type confidential info such as password, pin code, etc.,&\\
    &there are specific input types that allow hiding the &\\
    Use of clear text&information when typing, such as “textPassword”, &By developing malware,  an attacker can\\
    password field&“numberPassword”, etc. Instead of using these recommended&capture password typing by the target, \\
    (CWE-549)&input fields, many apps use clear text fields to enter this&or by observing the entry password.\\
    &kind of sensitive info. This allows an adversary to observe&\\
    &the entry password, to be copied in the clipboard, or to be&\\
    & screen captured using malicious apps.&\\
    \hline
    Sensitive data logged&Apps log sensitive data such as username and password.&Logged data can be accessed by privileged\\
    (CWE-532, CWE-534)&Those logged data will be stored in the logcat memory.&system apps, rooting or jailbreaking devices,\\
    &&or by developing mobile malware.\\
    \hline
   Hard-coded&For authentication to a backend server, the user must choose&By reverse-engineering the app, an attacker\\
    backend&his/her credentials (username, password). In some apps,&can get those credentials and steal sensitive\\
    credentials&backend credentials are hard-coded for connection to a&data from the backend resource.\\
    (CWE-798)&backend resource.&\\
  \hline
\end{tabular}
\end{adjustbox}
\end{table*}
\\

\begin{adjustbox}{width=\linewidth}
    \fcolorbox{black}{lightgray!40}{
        \begin{varwidth} {\linewidth}
            \textbf{RQ1:} WAEMU banking apps present numerous security issues. Most of them have been found in the developers’ code (instead of libraries). 
            Many security issues within the apps could be exploitable using various methods towards reverse engineering apps, developing malware, etc. 
        \end{varwidth}
    }
\end{adjustbox}


\subsection{Security issue evolution across the app versions}
This section describes the evolution of security issues through the various app versions.
Among the 59 MBAs considered in this study, 47 have at least two versions, including the reference one. Recall that not all apps have the same number of versions.
To investigate the evolution, we only consider 3 versions for each app among those with at least 3 versions, representing 68\% (32 over 47 MBAs). 

Table~\ref{tab:evol_1} illustrates the number of security issues on the chosen versions. 
Around 20 (62\%) out of the 32 MBAs have had increasing security issues between the first and intermediate versions. 
This number is around 22 (68\%) between the intermediate and the reference versions.
Globally, approximately 21 (representing 65\%) of these MBAs have had increasing security issues since the first version was available.
However, some of them have seen a decreased number of security issues on the intermediate version, but it has increased significantly in the reference one. 
\begin{table}[t!]
    \caption{Security issue evolution across the app versions.}
    \label{tab:evol_1}
    \begin{adjustbox}{width=0.6\linewidth,center}
        \begin{tabular}{l||ccc}
                &\textbf{\#SRCSs in}&\textbf{\#SRCSs in}&\textbf{\#SRCSs in}\\
                \textbf{Apps}&\textbf{1st version}&\textbf{intermediate version}&\textbf{reference version}\\
                & (oldest version) & & (latest version)\\
            \hline 
                A1&49&45&100\\
                
                \rowcolor{lightgray!50} A2&29&32&38\\
                
                A5&12&20&28\\
                
                \rowcolor{lightgray!50} A7&99&29&203\\
                
                A9&60&100&105\\
                
                \rowcolor{lightgray!50} A12&175&135&112\\
                
                A18&95&45&103\\
                
                \rowcolor{lightgray!50} A20&58&76&161\\
                
                A21&100&102&101\\
                
                \rowcolor{lightgray!50} A23&49&52&159\\
                
                A24&59&25&53\\
                
                \rowcolor{lightgray!50} A25&34&14&49\\
                
                A26&42&108&490\\
                
                \rowcolor{lightgray!50} A27&34&36&30\\
                
                A28&84&49&116\\
                
                \rowcolor{lightgray!50} A29&48&4&26\\
                
                A31&35&18&24\\
                
                \rowcolor{lightgray!50} A32&28&29&55\\
                
                A33&30&33&25\\
                
                \rowcolor{lightgray!50} A35&78&81&34\\
                
                A36&82&28&119\\
                
                \rowcolor{lightgray!50} A37&30&119&18\\
                
                A38&20&26&27\\
                
                \rowcolor{lightgray!50} A39&20&35&133\\
                
                A40&22&68&77\\
                
                \rowcolor{lightgray!50} A41&28&26&50\\
                
                A42&133&129&95\\
                
                \rowcolor{lightgray!50} A43&66&70&38\\
                
                A44&35&98&77\\
                
                \rowcolor{lightgray!50} A45&8&46&52\\
                
                A46&103&107&48\\
                
                \rowcolor{lightgray!50} A47&49&68&155\\
        \hline
        \end{tabular}
    \end{adjustbox}
\end{table}

To determine the percentage of security concerns that disappeared and the number of new ones, we considered all 47 apps. 
In Table~\ref{tab:evol_2}, we give the means of the number and percentage of SRCSs that disappeared and the number of new ones between the versions.
We can see that almost all the WAEMU banking apps have SRCSs disappeared, and many new ones appear from the first to the reference app.
\begin{table}[t!]
  \caption{Tracking of the security issues across the app versions.}
  \label{tab:evol_2}
  \begin{adjustbox}{width=0.6\linewidth,center}
  \begin{tabular}{l||ccc}
    &\textbf{Average of}&\textbf{Average of}&\textbf{Average of}\\
    \textbf{Apps}&\textbf{disappear SRCSs (\#)}&\textbf{disappear SRCSs (\%)}&\textbf{new SRCSs (\#)}\\
\hline
    A1&6.33&13.67&23.33\\
    \rowcolor{lightgray!50} A2&2.25&7.35&4.5\\
    A3&0&0&17\\
     \rowcolor{lightgray!50} A4&2&5&16\\
     A5&3.25&18.75&7.25\\
     \rowcolor{lightgray!50} A6&41&40.2&46\\
     A7&36.5&36.87&88.5\\
     \rowcolor{lightgray!50} A8&29&25.22&48\\
     
     A9&24.25&21.89&35.5\\
     
     \rowcolor{lightgray!50} A10&2&22.22&3\\
     
     A11&2&10&99\\
     
     \rowcolor{lightgray!50} A12&67.67&48.43&46.67\\
     
     A13&0&0&14\\
     
     \rowcolor{lightgray!50} A14&2&5.88&7\\
     
     A15&0&0&11\\
     
     \rowcolor{lightgray!50} A16&2&22.22&2\\
     
     A17&2&22.22&2\\
     
     \rowcolor{lightgray!50} A18&34&37.87&36\\
     
     A19&28&58.33&49\\
     
     \rowcolor{lightgray!50} A20&42&35.28&67.75\\
     
     A21&32&35.6&32.25\\
     
     \rowcolor{lightgray!50} A22&12&24&14\\
     
     A23&25&48.19&80\\
     
     \rowcolor{lightgray!50} A24&14.5&27.82&13\\
     
     A25&12.5&40.97&20\\
     
     \rowcolor{lightgray!50} A26&22.25&31.92&134.25\\
     
     A27&5.67&15.85&4.33\\
     
     \rowcolor{lightgray!50} A28&17.75&24.14&25.75\\
     
     A29&17.33&72.5&10\\
     
     \rowcolor{lightgray!50} A30&0&0&16\\
     
    A31&5.75&18.35&3\\
    
    \rowcolor{lightgray!50} A32&3.5&12.5&17\\
    
    A33&8.5&24.9&7.25\\
    
    \rowcolor{lightgray!50} A34&0&0&0\\
    
    A35&36.25&32.01&25.25\\
    
    \rowcolor{lightgray!50} A36&24.33&38.84&36.67\\
    
    A37&32&38.35&29\\
    
    \rowcolor{lightgray!50} A38&0.75&2.96&2.5\\
    
    A39&5&19.44&33.25\\
    
    \rowcolor{lightgray!50} A40&5.75&10.9&19.5\\
    
    A41&3&10.92&8.5\\
    
    \rowcolor{lightgray!50} A42&56.75&48.85&47.25\\
    
    A43&12&17.26&2.67\\
    
    \rowcolor{lightgray!50} A44&28.75&52.23&39.25\\
    
    A45&3.75&9.34&14.75\\
    
    \rowcolor{lightgray!50} A46&18.5&18.74&4.75\\
    
    A47&9&15.2&35.5\\
  \hline
\end{tabular}
\end{adjustbox}
\end{table}
\\

\begin{adjustbox}{width=\linewidth}
    \fcolorbox{black}{lightgray!40}{
        \begin{varwidth} {\linewidth}
             \textbf{RQ2:}
            We observe an increasing trend of security issues with the update of apps for most of the studied MBAs. Indeed, while developers propose updates, the new versions are not necessarily more secure: some issues are fixed while new ones are introduced in the process.
        \end{varwidth}
    }
\end{adjustbox}

\subsection{App security comparison}
\textbf{General comparison}.
\begin{figure}[t!]
  \centering
  \includegraphics[width=3.4in]{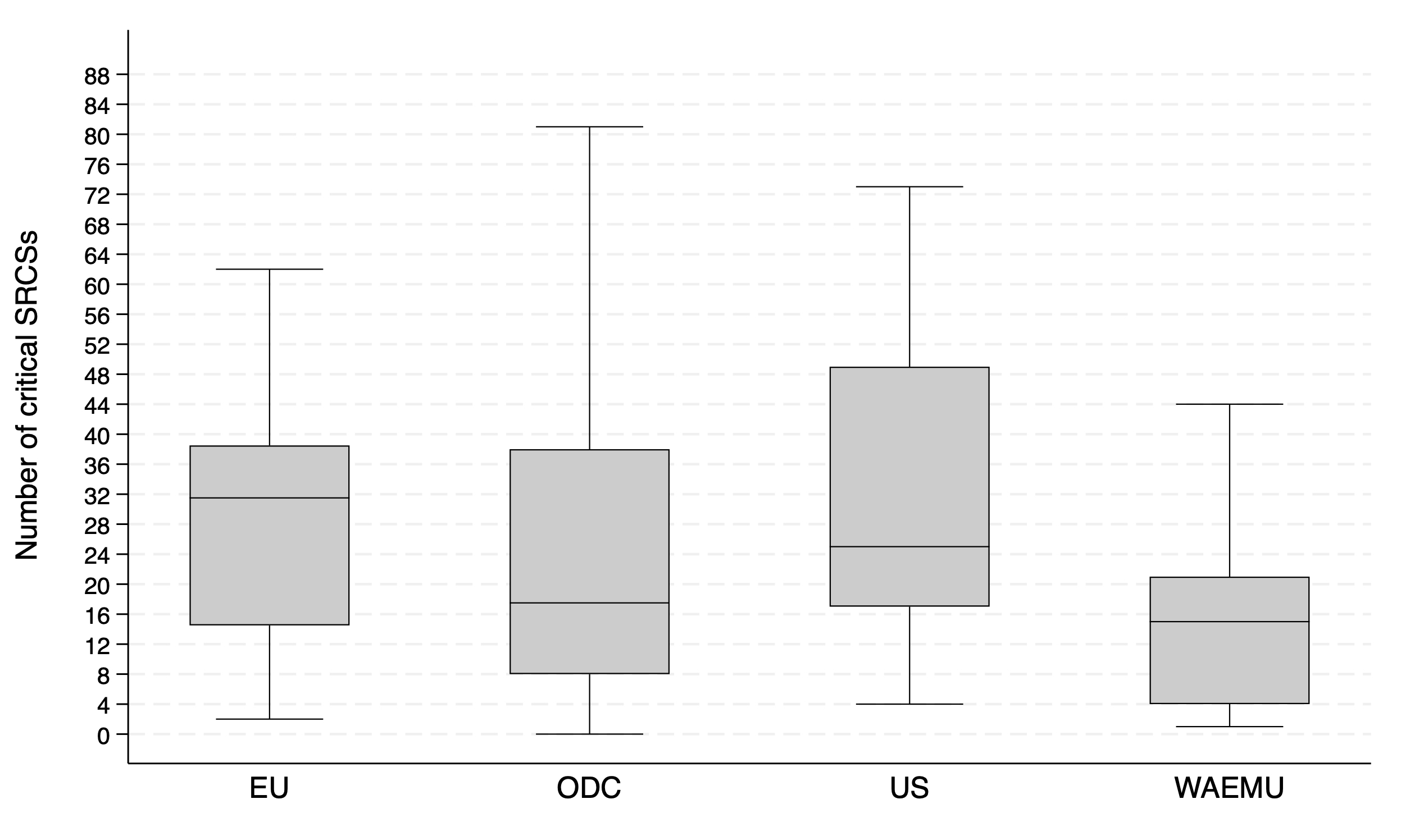}
  \caption{Comparison with WAEMU apps vs. EU, US, and other developing countries banking apps.}
  \label{fig:comparison1}
\end{figure}
This comparison is based on all critical issues found in the apps.
In other words, we take all the most critical issues from WAEMU banking apps and compare them with those coming from apps of the top 20 European Union (EU), top 20 United States (US), and top 20 other developing countries (ODC) banks.
The results show that WAEMU banking apps are more secure than those of EU, US, and ODC banks.
Indeed, as illustrated in Fig.~\ref{fig:comparison1}, 50\% of the WAEMU banking apps have at least 15 critical issues, compared to EU, US, and ODC apps, where 50\% of the apps have respectively more than 31, 25, and 17.
Moreover, an average of approximately 24 critical security issues per app is found in WAEMU apps.
EU, US, and ODC banking apps present approximately 31, 39, and 38 critical security issues per app, respectively. 
We have also compared the security of those apps based on more specific criteria, which we will explain next.

\begin{figure}[t]
  \centering
  \includegraphics[width=3.4in]{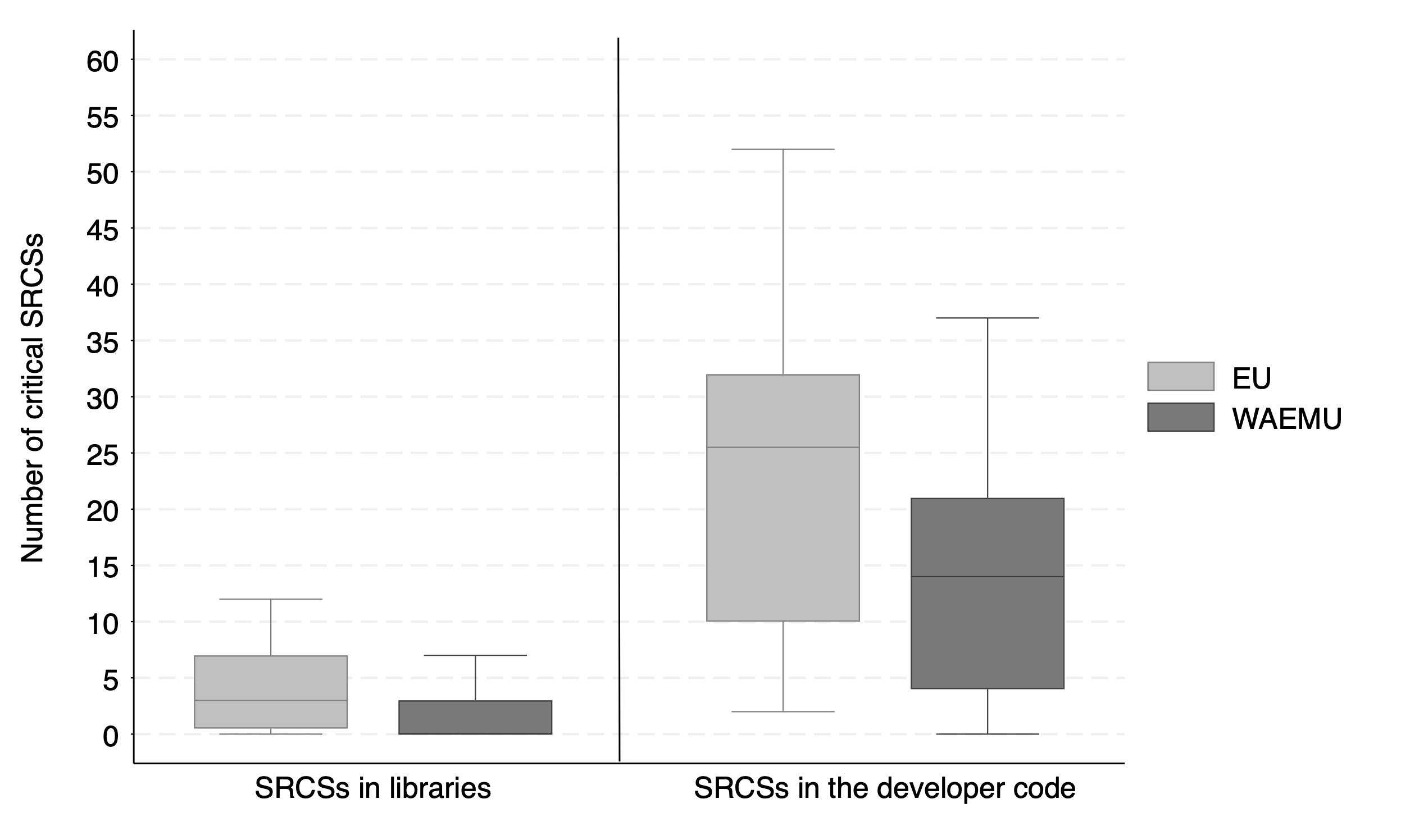}
  \caption{Comparison with WAEMU vs. EU banking apps based on the security issue location.}
  \label{fig:comparison2}
\end{figure}

\begin{figure}[t]
  \centering
  \includegraphics[width=3.4in]{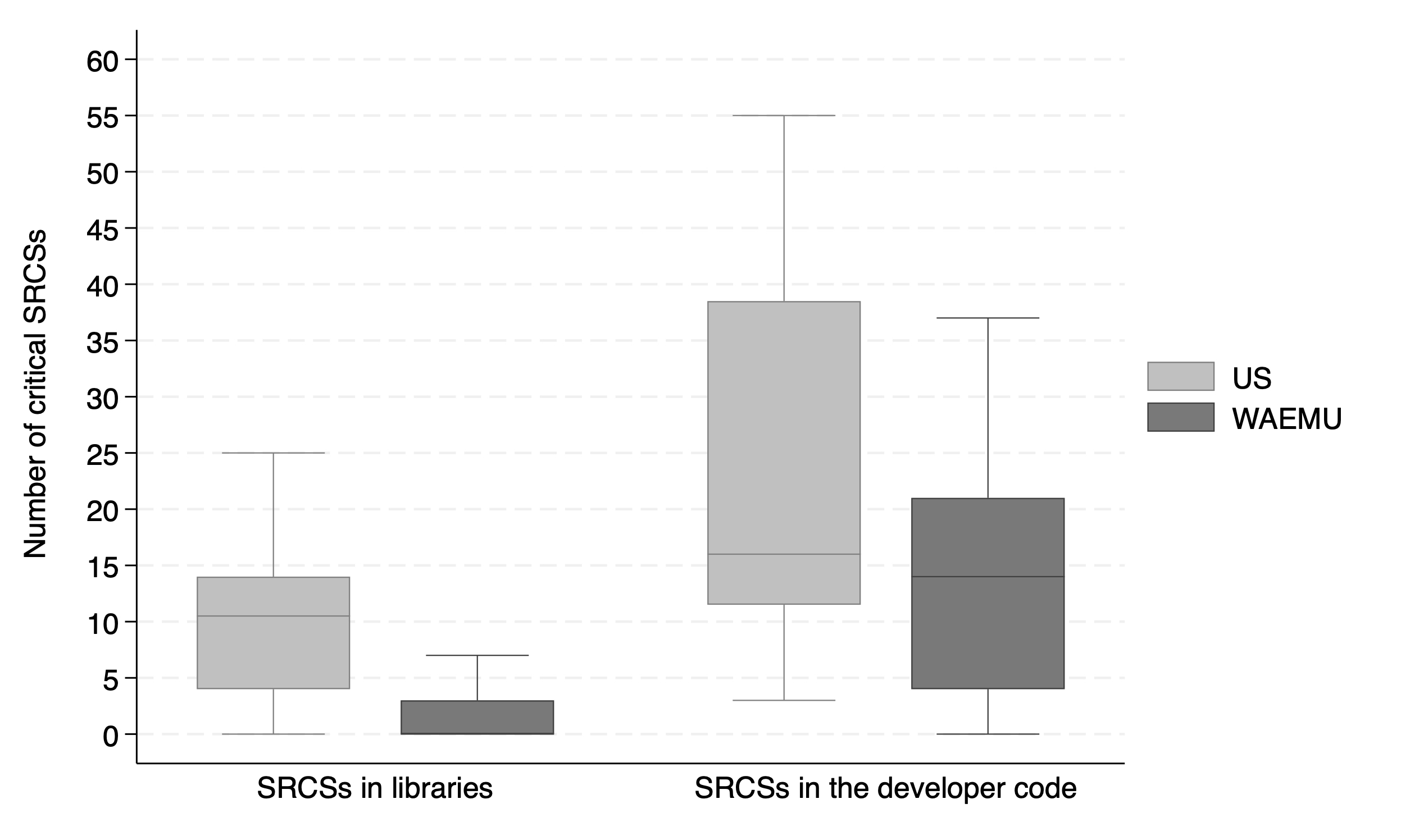}
  \caption{Comparison with WAEMU vs. US banking apps based on the security issue location.}
  \label{fig:comparison3}
\end{figure}

\textbf{Comparison based on the security issue location}.
We compare the security of the MBAs based on the location of the security issues found.
For EU banking apps compared to WAEMU banking apps, we found that around 75\% of the EU apps have at least one critical issue found in libraries used against 50\% of the apps from WAEMU, as illustrated in Fig.~\ref{fig:comparison2}. 
Furthermore, each of the EU apps has two or more critical issues in the developer code, unlike WAEMU apps, which have some that do not have any issues with the developer code.
As shown in Fig.~\ref{fig:comparison3}, US banking apps have more issues found in libraries and the developer code than WAEMU banking apps. 
Indeed, Fig.~\ref{fig:comparison3} shows that 50\% of the US apps have more than ten critical issues in libraries when only 50\% of the WAEMU apps have critical issues found in libraries, and the number does not reach ten.
In addition, each US banking app presents at least three critical issues in the developer code, whereas around 25\% of the WAEMU apps have less than four critical issues.
\begin{figure}[h]
  \centering
  \includegraphics[width=3.4in]{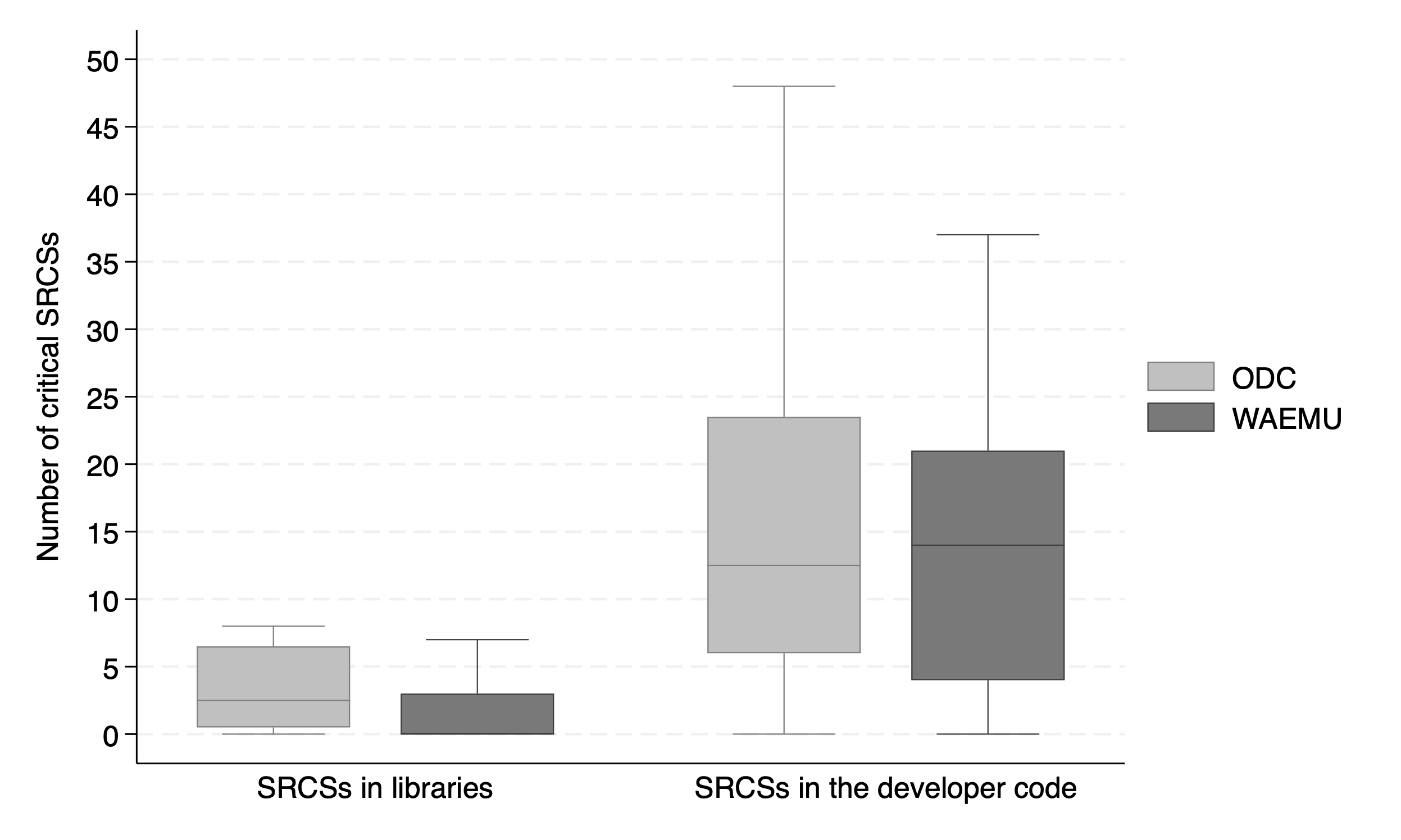}
  \caption{Comparison with WAEMU apps vs. other developing countries' banking apps based on the security issue location.}
  \label{fig:comparison4}
\end{figure}
As for MBAs from other developing countries, Fig.~\ref{fig:comparison4} shows that there are around 75\% of them present a critical issue found in libraries.
In addition to this, around 50\% of them have at least two critical issues, whereas only 50\% of WAEMU apps present at least one critical issue found in libraries.
Besides that, Fig.~\ref{fig:comparison4} also shows that 50\% of the MBAs from other developing countries present at least 12 critical issues in developer code compared to WAEMU apps, in which 50\% present 14 critical issues in the developer code.

\textbf{Comparison based on the top ten most common critical issues}. 
\begin{figure*}
     \centering
     \begin{subfigure}[b]{0.4\textwidth}
         \centering
         \includegraphics[width=1.1\textwidth]{Top_10_vulnerabilities.jpg}
         \subcaption{Top ten of WAEMU apps' most common critical issues.}
         \label{fig:topWAEMU}
     \end{subfigure}
     \hfill
     \begin{subfigure}[b]{0.4\textwidth}
         \centering
         \includegraphics[width=1.1\textwidth]{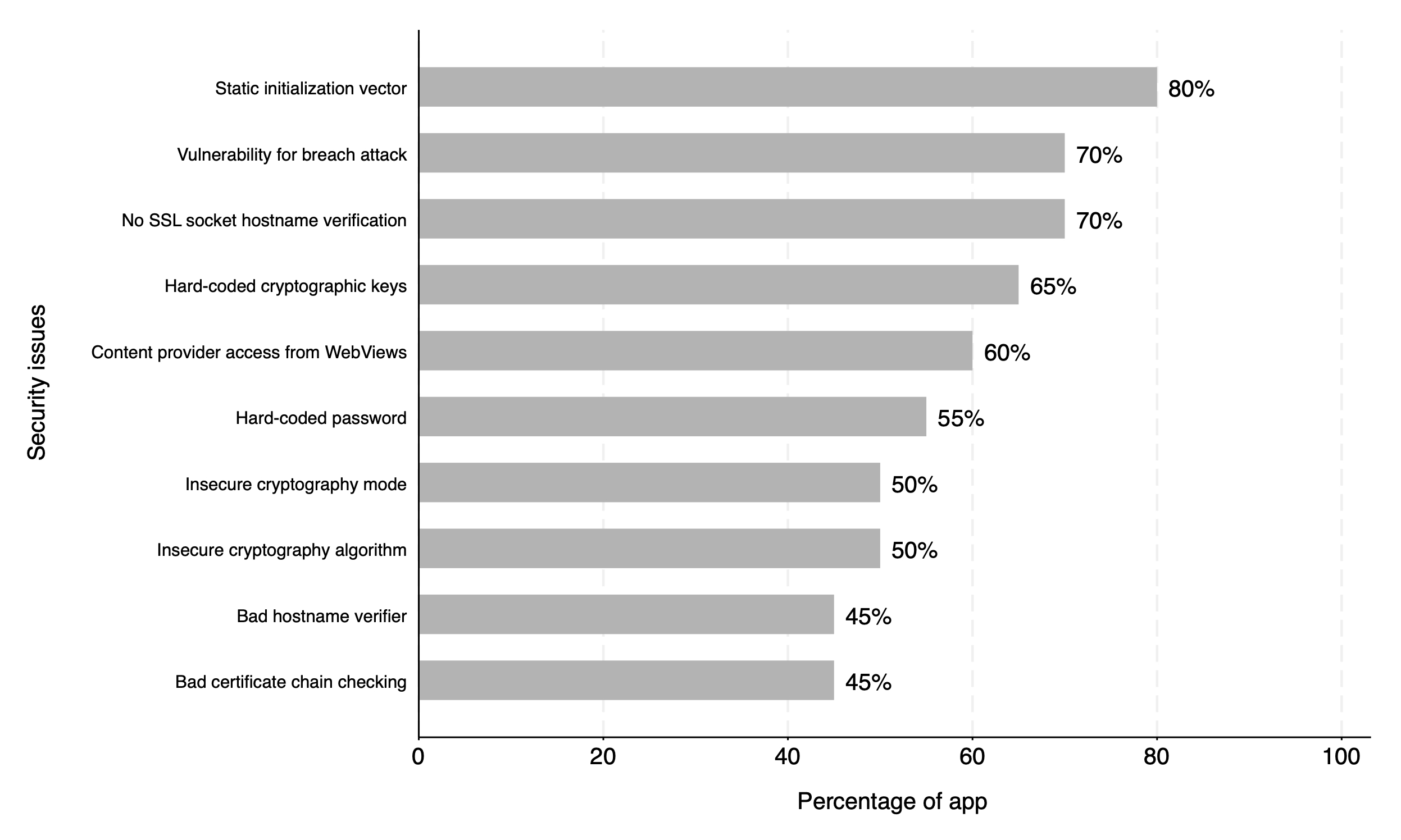}
         \subcaption{Top ten of EU apps' most common critical issues.}
         \label{fig:topEU}
     \end{subfigure}
    \hfill
     \begin{subfigure}[b]{0.4\textwidth}
         \centering
         \includegraphics[width=1.1\textwidth]{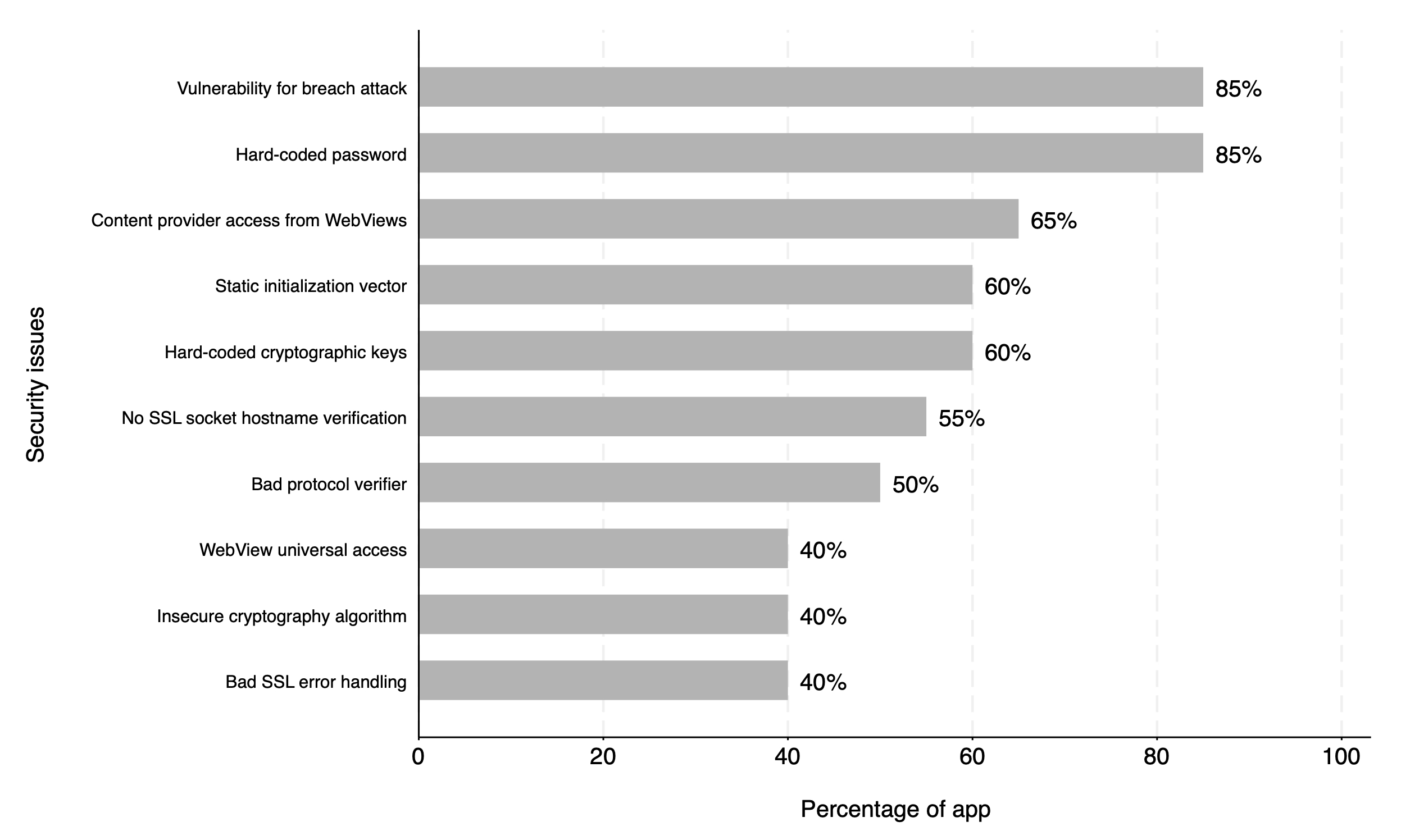}
         \subcaption{Top ten of US apps' most common critical issues.}
         \label{fig:topUS}
     \end{subfigure}
    \hfill
     \begin{subfigure}[b]{0.4\textwidth}
         \centering
         \includegraphics[width=1.1\textwidth]{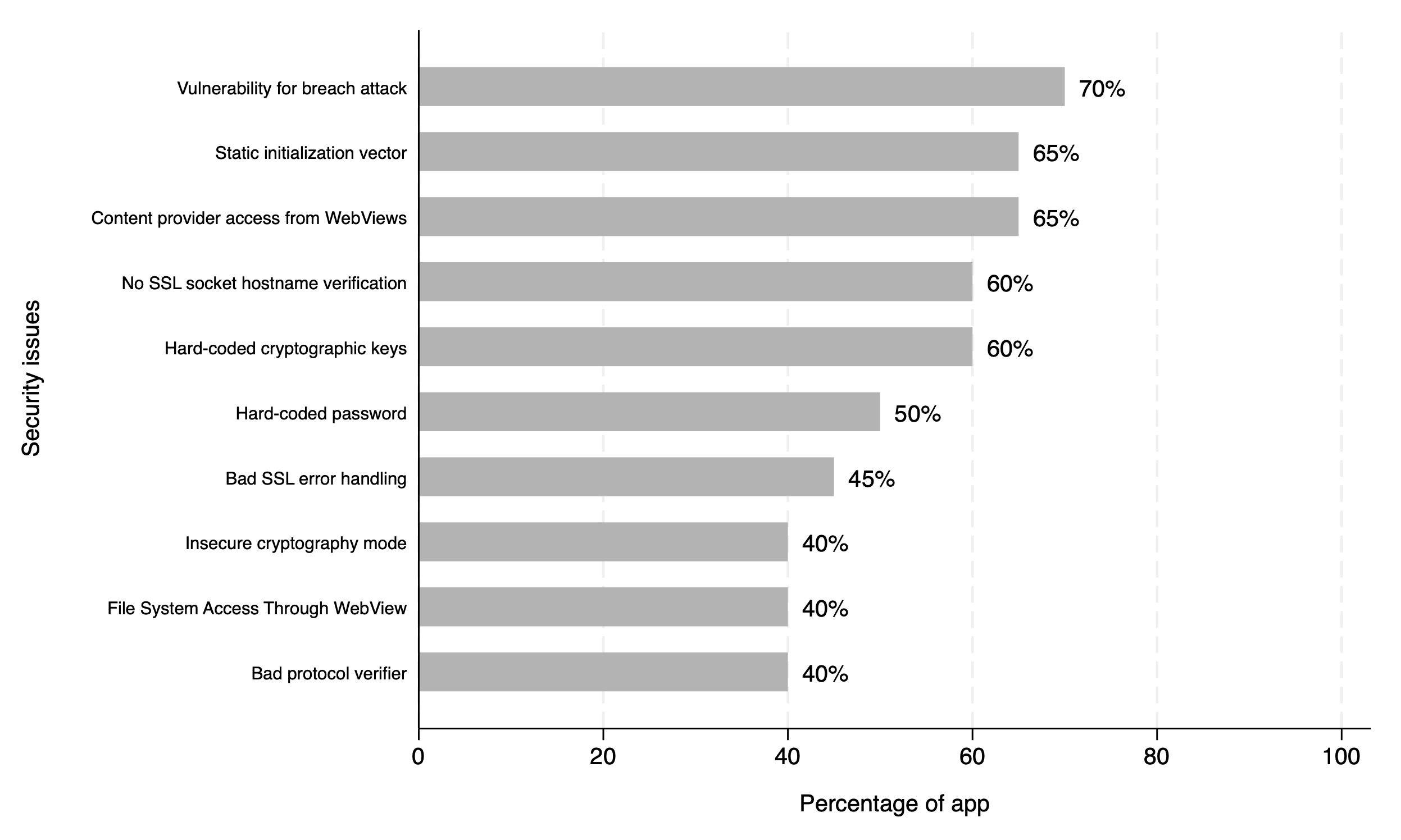}
         \subcaption{Top ten of ODC apps' most common critical issues.}
         \label{fig:topODC}
     \end{subfigure}
     \caption{Comparison of WAEMU and EU, US, and other developing country banking apps.}
     \label{fig:App_comp}
\end{figure*}
Most of the top security issues in EU, US, and ODC are the same in WAEMU, as highlighted in Fig.~\ref{fig:App_comp}, including static initialization vector (IV), hard-coded password, bad SSL error handling and file system access through WebView.

A random initialization vector (IV) ensures that encrypting the same data multiple times leads to different ciphertexts. This precaution prevents attackers from matching ciphertexts against pre-computed tables (e.g., Rainbow Tables) to derive the plaintext. If the IV is constant, this protection is rendered ineffective, and tables can be pre-computed for this IV. This problem is common in EU, US, and ODC banking apps as illustrated in Fig.~\ref{fig:topEU},~\ref{fig:topUS}, and~\ref{fig:topODC}.

In WAEMU apps, on the other hand, most apps use old algorithms for encrypting and decrypting data, as highlighted by the high number of apps containing insecure cryptography algorithm issues in Fig.~\ref{fig:topWAEMU}. Such algorithms can be broken regardless of a potential IV problem.

At least half of the EU, US, and ODC banking apps use hard-coded passwords to derive a cryptography key, which is then used to encrypt and decrypt data. Such behavior is more complex than directly hard-coding the key but leads to the same vulnerability. In 20\% of the WAEMU apps, this key is directly hard-coded, avoiding the key derivation step. In either case, attackers can reconstruct the key and decrypt the data.

An SSL error can occur when a client loads a resource through webView.
Consequently, the applications should properly handle SSL errors, allowing SSL connections to be aborted when errors occur.
In US and ODC apps, as presented by their top ten in Fig.~\ref{fig:topUS} and~\ref{fig:topODC}, many apps ignore SSL certificate errors, making them vulnerable to man-in-the-middle attacks. WAEMU apps, on the other hand, do not have these issues in their top ten. 

As highlighted by the percentage in Fig.~\ref{fig:topODC}, many ODC apps configure a WebView such that the WebView can only access files on the device's file system. In WAEMU apps, this practice is uncommon. However, they use universal access, which allows access to any resource, including files.
\\

\begin{adjustbox}{width=\linewidth}
    \fcolorbox{black}{lightgray!40}{
        \begin{varwidth} {\linewidth}
            \textbf{RQ3:} 
            Banking apps from the WAEMU appear to present fewer critical issues than MBAs from the top 20 of the EU, US, and other developing countries (ODC). 
            Their security issues, as well as those of the apps from the EU, the US, and other developing countries, come from libraries and developers' code. However, WAEMU apps present less critical issues in the developer code, as well as in libraries.
            Based on the top 10 security issues, WAEMU apps have avoided some of the critical issues that others have, but some specific issues are more prevalent.
        \end{varwidth}
    }
\end{adjustbox}

\subsection{Child and parent banking apps}
Some of the WAEMU financial institutions are subsidiaries (child banks) of banks from other regions (parent banks). Some parent banks are from African countries (African parent banks), and others are from non-African countries (International parent banks). In this section, we perform a security app comparison of child banking apps with their respective parent banking apps from both regions.

On the one hand, 13 over 59 MBAs are African child banking apps.
As illustrated in Table~\ref{tab:african_chlid}, all of them inherit security issues from African parent banking apps, representing a mean of 37\% security issues inherited.
Furthermore, at least 17\% of their security issues are new ones.

On the other hand, 6 over 59 MBAs are International child banking apps.
As illustrated in Table~\ref{tab:international_chlid}, all of the International child banking apps inherit security issues from International parent banking apps, representing a mean of 20\% security issues inherited.
More than 65\% of their security issues are new. 
\begin{table}[t!]
  \caption{Comparison of SRCS numbers between African child and parent apps.}
  \label{tab:african_chlid}
  \begin{adjustbox}{width=\linewidth,center}
  \begin{tabular}{c||lcccc}
    \textbf{African Parent}&\textbf{Child}&\textbf{\# of child}&\textbf{\# of parent }&\textbf{\% of inherit}&\textbf{\% of new}\\
    \textbf{banking apps (AP)}&\textbf{apps}&\textbf{app SRCSs}&\textbf{app SRCSs}&\textbf{ SRCSs}&\textbf{SRCSs}\\
    \hline
    &A1&100&33&22\%&78\%\\
    AP1&A5&28&33&54\%&46\%\\
     &A15&28&33&54\%&46\%\\
    \hline
    AP2&A11&117&38&11\%&89\%\\
    \hline
     &A14&39&72&33\%&67\%\\
    &A24&53&72&30\%&70\%\\
    AP3&A25&49&72&43\%&57\%\\
     &A41&50&72&44\%&56\%\\
    &A47&155&72&14\%&86\%\\
    &A48&23&72&83\%&17\%\\
     \hline
     AP4&A29&26&36&54\%&46\%\\
     \hline
     AP5&A32&55&18&18\%&82\%\\
     \hline
     AP6&A38&27&35&19\%&81\%\\
  \hline
\end{tabular}
\end{adjustbox}
\end{table}
\begin{table}[h!]
  \caption{Comparison of SRCS numbers between International child apps and International parent apps.}
  \label{tab:international_chlid}
  \begin{adjustbox}{width=\linewidth,center}
  \begin{tabular}{c||lcccc}
    \textbf{Intrenational Parent}&\textbf{Child}&\textbf{\# of child}&\textbf{\# of parent }&\textbf{\% of inherit}&\textbf{\% of new}\\
    \textbf{banking apps (IP)}&\textbf{apps}&\textbf{app SRCSs}&\textbf{app SRCSs}&\textbf{ SRCSs}&\textbf{SRCSs}\\
    \hline
    &A16&29&42&14\%&86\%\\
    IP1&A17&29&42&14\%&86\%\\
    &A49&12&42&33\%&67\%\\
    \hline
    IP2&A36&119&29&10\%&90\%\\
    \hline
    IP3&A39&133&53&26\%&74\%\\
     \hline
    IP4&A40&77&48&21\%&79\%\\
  \hline
\end{tabular}
\end{adjustbox}
\end{table}
\\

\begin{adjustbox}{width=\linewidth}
    \fcolorbox{black}{lightgray!40}{
        \begin{varwidth} {\linewidth}
            \textbf{RQ4:} Child apps always inherit security issues from the parent apps, with an important number. Further, they always present several new ones.
        \end{varwidth}
    }
\end{adjustbox}

\section{Discussions and Recommendations}
\label{discuss}
\subsection{Discussion about results}


Our analysis also reveals many SRCSs on the developers' code and the libraries used by the WAEMU banking apps.
However, most of them have been found in the developers' code, suggesting that developers compromise, intentionally or not, the security of users. 
To avoid such security issues, developers must understand the best practices and know how the required features of a banking app can be implemented securely.

Based on the most critical security issues (csi) found in the MBAs, WAEMU banking apps present fewer issues (average of 24 csi/app approximatively) than EU, US, and ODC banking apps, which present 31, 39, and 38 csi/app respectively.
Most of these security issues have been found in the developer code.
On average, WAEMU apps have fewer issues than EU, US, and ODC banking apps introduced by developers.
This could be explained by the higher number of lines of code in non-WAEMU banking applications than in WAEMU banking applications.
It could simply mean that WAEMU banking application developers are more concerned with best practices when developing applications than non-WAEMU banking application developers as well.

Among the security issues found, many could be potentially exploitable.
For instance, cryptography is used to secure users' sensitive data and avoid unauthorized access by attackers~\cite{owasp1}. However, many cryptographic algorithms are considered outdated and insecure over time, including SHA1 and MD5. However, they are still used prevalently in mobile banking apps. As illustrated in Listing 1, apps often use insecure hash functions to hash sensitive data.
In this app, when updating the password, the new one is recuperated (\textit{in line 4}), hashed using md5, and put in a string variable (\textit{in line 32}).
When a malicious person retrieves this string value, the password can be easily recovered.
This practice is the most common since it affects 80\% of the WAEMU apps.
This suggests that all the affected apps could be potentially compromised, leading to significant impacts, including data breaches, data loss, and reputation damage~\cite{guardrails1}.
It is also important to use strong cryptographic algorithms. Still, it is also important for developers to check the best practices regularly since algorithms considered more secure could be insecure over time.
Besides that, banking apps often use a hard-coded cryptographic key to encrypt and decrypt data. If this key is known by a malicious person (which is trivial to achieve by decompiling the app), it significantly increases the possibility of recovering the encrypted data~\cite{mappings2010cwe}. The code snippet in Listing 2 shows an example of such a hard-coded key (\textit{line 4}) being used for generating a secret key (\textit{line 9}) which is used in the \textbf{encode} (\textit{line 12}) and \textbf{decrypt} (\textit{line 19}) methods.
These methods are later used for encrypting and decrypting data, including passwords, one-time passwords (OTP), and secret questions/answers.
Our analysis found this in 20\% of the apps, representing the $10^{th}$ most common security issues.
There are more vulnerabilities among the ten most prevalent security issues such as access to content providers within WebView (78\%), which may allow access to protected content~\cite{codeQL}, bad hostname verifier (51\%), which may allow certificates with incorrect or mismatched hostnames to be accepted leading to data breaches, bad certificate chain checking (47\%), which could potentially trust a certificate that has not been issued by a trusted certificate authority or that has been tampered with allowing a man-in-the-middle attack, no SSL socket hostname verification (32\%) when establishing connections, which can lead to a man-in-the-middle attack, and more.
Several MBAs have the same developer, which could explain the frequency of certain security problems.

Even outside the top 10 list, there are other severe vulnerabilities that, albeit not very common, can be potentially exploitable and affect several apps. For example, several WAEMU apps log sensitive information such as the user's password under some circumstances, as illustrated at \textit{line 25} in listing 3.
This vulnerability could be exploitable by simply connecting the device to a PC and using adb\footnote{\url{https://developer.android.com/tools/adb}} commands~\cite{svoboda}.
Other apps installed in the same device, such as malicious or privileged system apps, could also potentially exploit this vulnerability.

Developers should avoid bad practices, including hard-coded keys and credentials and logging sensitive data.
They must properly verify the hostname and certificate authority before connecting with servers through the apps. 
\begin{lstlisting}[language=Java, caption=App using an insecure crypto algorithm.]
public class classeName {
    [...]
    final EditText editText = (EditText) findViewById(R.id.txt0ldPassword)Al
    final EditText editText2 = (EditText) findViewById(R.id.txtNewPassword) ;
    final Editrext editiexts = (Edittext) rindviewbyid(K.d.txtcontirmpassword);
    ((Button) findViewById(R.id.btnChange)).setOnClickListener(new View.OnClickListener) { 
    @Override 
    public void onClick(View view) {
        final String str2;
        if (...) {
        [...]
        } 
        else {
            String obj = editText2.getText(). toString();
            try {
                MessageDigest instance = MessageDigest.get Instance("MD5");
                instance.update("P@#$&*(-+"-getBytes(), 0, 10);
                String str3 = obj * new Biginteger(1, instance.digest).toString(16);
                MessageDigest instance2 = MessageDigest.getInstance("MD5");
                instance2.update(str3.getBytes(), 0, str3.length());
                obj = new BigInteger(1, instance2.digest()).toString(16);
            } catch (Exception unused) {
            }
            [...]
            String obj2 = editText2.getText).toString();
            try {
                MessageDigest instance3 = MessageDigest-getInstance("MD5");
                instance3.update("P@#$&*(-+"-getBytes(), 0, 10);
                obj2 = obj2 + new BigInteger(1, instance3.digest()).toString(16);
                MessageDigest instance4 = MessageDigest.getinstance("MD5");
                instance4.update(obj2.getBytes(), 0, obj2. length()):
                str2 = new BigInteger(1, instance4.digest()).toString(16);
            } catch (Exception unused2) {
                    str2 = obj2;
            }  
            [...]
        }
        [...]
    }
    [...]
}
\end{lstlisting}
\clearpage
\begin{lstlisting}[language=Java, caption=App using a hard-coded key.]
public class globalmethods {
    private static final String ALGO = "AES";
   [...]
    private static final byte[] keyValue = {64, 97, 112, 101, 120, 53, 48, 102, 116, 119, 64, 114, 51, 53, 49, 56};
    
    [...]
    
    private static Key generateKey() throws Exception {
        return new SecretKeySpec(keyValue, ALGO);
    }

    public static String encode(String str) throws Exception {
        Key generateKey = generateKey();
        Cipher instance = Cipher.getInstance(ALGO);
        instance.init(1, generateKey);
        return Base64.encodeToString(instance.doFinal(str.getBytes()), 0).replace(IOUtils.LINE_SEPARATOR_UNIX, "");
    }

    public String decrypt(String str) {
        try {
            Key generateKey = generateKey();
            Cipher instance = Cipher.getInstance(ALGO);
            instance.init(2, generateKey);
            return new String(instance.doFinal(Base64.decode(str.getBytes(), 0)));
        } catch (Exception unused) {
            return str;
        }
    }
\end{lstlisting}

    
    


\begin{lstlisting}[language=Java, caption=App logging a password.][t!]
public class CodeSecretActivity extends AppCompatActivity {
    private Button btnConfirmer;
    private EditText editTextCodeSecret, editTextNewCodeSecret;
    private Intent passwordIntent;
    private MyProgressDialog pbar;
    @Override
    public void onCreate(Bundle bundle) {
        new ProgressUpdatePassword().execute(stringExtra, stringExtra2, CodeSecretActivity.this.editTextNewCodeSecret.getText().toString());
        this.editTextCodeSecret = (EditText) findViewById(R.id.edittext_code_login);
        this.editTextNewCodeSecret = (EditText) findViewById(R.id.edittext_new_code_login);
        this.btnConfirmer = (Button) findViewById(R.id.btn_new_code);
        Intent intent = getIntent();
        this.passwordIntent = intent;
        final String stringExtra = intent.getStringExtra("username");
        final String stringExtra2 = this.passwordIntent.getStringExtra("password");
        this.btnConfirmer.setOnClickListener(new View.OnClickListener() { 
            @Override 
            public void onClick(View view) {
                if (...) { [...]
                }
                else {
                    new ProgressUpdatePassword().execute(stringExtra, stringExtra2, CodeSecretActivity.this.editTextNewCodeSecret.getText() .toString());
                    Log.i("username ", stringExtra);
                    Log.i("password ", stringExtra2);
                    Log.i("new password ", CodeSecretActivity.this.editTextNewCodeSecret.getText() .toString());}}
\end{lstlisting}

By investigating security issue evolution through the app versions, we identified an increased number between the first and the intermediate versions in most WAEMU apps (approximately 62\%), as well as between the intermediate and the reference versions (approximately 68\%).
This indicates that even if banks and financial institutions regularly propose updates (at least 3 versions in 68\% apps), these updates are ineffective in fixing the security issues. 
Ideally, updates should improve the security of the apps by fixing security issues.
Based on the numbers presented in Table~\ref{tab:evol_2}, several SRCSs have disappeared between the first to the last (reference) version.
However, the number of new SRCSs has risen significantly, almost doubling the number of disappeared SRCSs in most WAEMU apps.

We also noticed that several banks and financial institutions do not propose updates regularly.
For instance, in more than one year, they could propose a single update or not. 
This could suggest they do not care about their customers' security.
Apps should be regularly updated to consider new security trends, and issues must be fixed after updates.
\\

Of the 59 WAEMU banking apps, 19 out of them are from banks with parents in a non-WAEMU country, among which 13 have African parents (AP), while 6 have international parents (IP).
We noticed interesting findings by comparing these apps with those from the parent banks.
Indeed, the 19 (child banking apps) inherit security issues from parent banking apps: from 11 to 83\% of the issues for AP and from 14 to 33\% for IP.
It has also been noticed that even if all the child banking apps introduce several new security issues, most parent banking apps have more security issues than them.

\subsection{Limitations and future directions}

This work, investigating the security of mobile banking apps in WAEMU countries, highlights the security issues mostly introduced by the developers. 
In this study, we point out potentially exploitable security issues and propose use cases that could exploit them.
As it is important to know if these security issues are exploitable to better understand the security impacts, the future direction could be to propose approaches and methods for exploiting these issues.
When comparing the security of WAEMU apps with those from other regions, we focused solely on the total number of security issues, the ten most common issues, and their locations. However, relying on the number of security issues alone can introduce bias, as the number of issues may correlate with the app’s size or the number of lines of code. To address this, future investigations could consider the number of lines of code in each app to achieve more accurate and unbiased results.
Mobile money apps are as important as mobile baking apps since they contribute to financial inclusion.
Future research could investigate those apps and extend the study context rather than limiting it to WAEMU countries only.

\section{Threat to the validity}
\label{threat}
We gathered the latest app versions from WAEMU banks in December 2022. It is likely that from then until the time of this paper’s composition, banks have introduced new APK versions by addressing and rectifying security issues. This could potentially impact the relevance of this study’s findings.

We have conducted a manual review of the app source codes to verify the identified SRCSs and determine which ones pose vulnerabilities. However, due to time limitations and the abundance of certain SRCSs, we were unable to examine all SRCSs in every MBA. It is crucial to understand that an SRCS might be a vulnerability in some apps but not in others. For instance, a scanner might detect SHA1 or MD5 hashing sensitive data such as usernames and passwords in one app, while in another app, it might hash data that does not significantly compromise user security. 

We examined the evolution of security issues across app versions, focusing only on those with at least three versions. We tracked the occurrence of each SRCS within these apps, comparing its frequency between consecutive versions. If an SRCS’s frequency increased in the next version, we considered it as ‘increased’, with the increase being the difference in occurrence numbers between the two versions. If not, we considered the SRCS as ‘fixed’. 
Thus, we considered an SRCS ‘fixed’ if its occurrence number decreased in the subsequent version. 
It is important to note that an issue might disappear due to intentional fixing or code snippet deletion.  
An SRCS instance might be present in one version and disappear in the next, regardless of whether its occurrence number has surged. Conversely, new SRCSs might emerge even if the occurrence number has dropped. Given the difficulty in tracking disappeared SRCS instances, we focused solely on occurrence numbers. 

\section{Related Work}
\label{related}
To our knowledge, no study focused on mobile banking apps in WAEMU countries. 
However, several works have been performed on mobile banking and money apps in other contexts.

Bassolé et al.~\cite{Bassole} investigate Android banking and payment apps in African countries by identifying specific vulnerabilities and raising awareness of the security in those apps. The authors provide valuable information to developers and stakeholders, enabling them to improve security measures and protect user data. 

Bowers et al.~\cite{10.1145/3317549.3319723} perform a security analysis of digital credit apps in developing countries. In their study, they investigated the privacy policies of several companies and found previously undisclosed data types were collected. They investigated the configurations between apps and servers, as well, and they discovered widespread misconfiguration of encryption. 

In their work, Latifa et al.~\cite{10.1145/3108421.3108433} investigate the impacts that permissions could have on the security of users' mobile banking apps. By analyzing the permissions requested by some mobile banking apps from Maghreb countries, they found many apps use unnecessary and dangerous permissions. 

Bojjagani and Sastry~\cite{bojjagani2017vaptai} propose a threat model (VAPTAi) for enhancing the security of Android and iOS mobile banking apps.  
VAPTAi is designed for the assessment of vulnerabilities and penetration testing of mobile banking apps. 
This work is a bit similar to one of their prior work in which the same authors propose a tool that identifies threats at different levels, including app, network, and device levels~\cite{10.1007/978-3-319-28658-7_57}. 
In this work, they analyze vulnerabilities using techniques such as static, dynamic, and forensic analysis and identify several attack surfaces. 

Kaka et al.~\cite{7947811} evaluate the vulnerability of mobile banking apps from India by mainly performing man-in-the-middle (MiTM) attacks. 
In their work, the authors discover that in most apps, MiTM attacks could be successfully achieved even if the apps use HTTPS for communication with the servers.  

Castle et al.~\cite{10.1145/3001913.3001919} evaluate the security challenges of mobile money systems in the developing world by identifying vulnerabilities, assessing the factors contributing to them, and proposing potential solutions to enhance the security of such systems. In the same paper, the authors have given a response to a prior work in which Reaves et al.~\cite{reaves2015mo} identify and document the security and privacy issues in branchless banking apps in order to raise awareness among developers, financial institutions, and regulators about the importance of robust security measures. 

\section{Conclusion}
\label{concl}
This study performed a security assessment of mobile apps from banks and financial institutions in the West African Economic and Monetary Union (WAEMU) countries.
We have statically analyzed fifty-nine (59) collected mobile banking apps (MBAs) from the eight WAEMU countries, and the results show that there are many vulnerabilities that can lead to several real-world attacks.

An attacker could reverse-engineer the apps to retrieve hard-coded backend credentials and cryptographic key used for encrypting/decrypting data. Additionally, attackers could use dictionary attacks to crack hashes obtained from insecure algorithms such as SHA1 and MD5. They could develop mobile malware to extract sensitive data, such as usernames and passwords, from log files in the device or capture credentials entered into unprotected (clear text) UI fields. Most security issues are found in the developers’ code. 
This indicates that developers, whether intentionally or not, introduce vulnerabilities into mobile apps, resulting in user privacy violations and potential damages. As a result, the reputation of financial institutions can be compromised.

Some banks and financial institutions offer updates to their mobile banking apps (MBAs), but these updates often do not address all security issues. Additionally, new security vulnerabilities frequently emerge in the latest versions. Other institutions either do not provide updates or do so irregularly. The emergence of new threats and exploitation techniques can compromise app security if the latest protection techniques are not implemented and newly discovered security breaches are not addressed.

According to the results of the comparison of WAEMU apps with European Union (EU) apps, United States (US) apps, and developing countries other than WAEMU countries (ODC), MBAs from WAEMU seem to be more secure than those.
Some WAEMU apps are banking apps of subsidiaries of banks in other regions.
Our analysis shows that those apps always inherit some security issues of their parent apps, and almost all present several new ones.
This indicates that subsidiary apps seem to be more vulnerable and can compromise user security much more than parent apps.

As part of responsible disclosure, we have contacted some banks to discuss our findings regarding their MBAs.
The goal was to help them fix these concerns to improve the security of their apps.
It was hard to find a security contact for reporting in many banks. 
In some cases, the email address in the Play Store did not exist or was a Gmail address.
We did not get responses from the banks we tried to contact. 
No bank had a proper vulnerability disclosure process.

We have given recommendations for the developers to allow them to avoid some practices that can lead to damage and adopt more secure practices.

\section{Data Availability}
\label{sec:data_availability}
To promote transparency in scientific research, we make all of the artifacts used in this study available to the community:

\begin{center}
    \url{https://github.com/liounea/Data_for_WAEMU_Apps_Papers}
\end{center}

\section*{Acknowledgments}

This work is supported by the Luxembourg Ministry of Foreign and European Affairs through their Digital4Development (D4D) portfolio under the project LuxWAyS (Luxembourg/West-Africa Lab for Higher Education Capacity Building in CyberSecurity and Emerging Topics in ICT4Dev.)

\bibliographystyle{IEEEtran}
\bibliography{bib}

\begin{thebibliography}{10}
\providecommand{\url}[1]{#1}
\csname url@samestyle\endcsname
\providecommand{\newblock}{\relax}
\providecommand{\bibinfo}[2]{#2}
\providecommand{\BIBentrySTDinterwordspacing}{\spaceskip=0pt\relax}
\providecommand{\BIBentryALTinterwordstretchfactor}{4}
\providecommand{\BIBentryALTinterwordspacing}{\spaceskip=\fontdimen2\font plus
\BIBentryALTinterwordstretchfactor\fontdimen3\font minus \fontdimen4\font\relax}
\providecommand{\BIBforeignlanguage}[2]{{%
\expandafter\ifx\csname l@#1\endcsname\relax
\typeout{** WARNING: IEEEtran.bst: No hyphenation pattern has been}%
\typeout{** loaded for the language `#1'. Using the pattern for}%
\typeout{** the default language instead.}%
\else
\language=\csname l@#1\endcsname
\fi
#2}}
\providecommand{\BIBdecl}{\relax}
\BIBdecl

\bibitem{bei}
\BIBentryALTinterwordspacing
B.E.I. (2016) Le secteur bancaire en afrique subsaharienne: Évolutions récentes et inclusion financière numérique. [Online]. Available: \url{https://www.eib.org/attachments/efs/economic_report_banking_africa_digital_financial_inclusion_fr.pdf}
\BIBentrySTDinterwordspacing

\bibitem{ashTurner}
\BIBentryALTinterwordspacing
A.~Turner. (2024) How many people have smartphones in the world? [Online]. Available: \url{https://www.bankmycell.com/blog/how-many-phones-are-in-the-world#}
\BIBentrySTDinterwordspacing

\bibitem{osiris}
\BIBentryALTinterwordspacing
Osiris. (2023) En afrique subsaharienne, le taux d’adoption des smartphones atteindra 87\% en 2030. [Online]. Available: \url{http://www.osiris.sn/En-Afrique-subsaharienne-le-taux-d.html}
\BIBentrySTDinterwordspacing

\bibitem{appRoov}
\BIBentryALTinterwordspacing
Approov. (2023) Security challenges of financial mobile apps in africa. [Online]. Available: \url{https://approov.io/info/security-challenges-of-financial-mobile-apps-in-africa}
\BIBentrySTDinterwordspacing

\bibitem{diallo2024security}
A.~Diallo, J.~Samhi, T.~Bissyand{\'e}, and J.~Klein, ``(in) security of mobile apps in developing countries: A systematic literature review,'' \emph{arXiv preprint arXiv:2405.05117}, 2024.

\bibitem{Ansong}
E.~D. Ansong and T.~Q. Synaepa-Addision, ``A comparative study of user data security and privacy in native and cross platform android mobile banking applications,'' in \emph{2019 International Conference on Cyber Security and Internet of Things (ICSIoT)}, 2019, pp. 5--10.

\bibitem{Uduimoh}
A.~Uduimoh, I.~Idris, O.~Osho, and S.~Abdulhamid, ``Forensic analysis of mobile banking applications in nigeria,'' \emph{i-manager's Journal on Mobile Applications and Technologies}, vol.~6, pp. 9--20, 06 2019.

\bibitem{Osho}
O.~Osho, U.~L. Mohammed, N.~N. Nimzing, A.~A. Uduimoh, and S.~Misra, ``Forensic analysis of mobile banking apps,'' in \emph{Computational Science and Its Applications -- ICCSA 2019}, S.~Misra, O.~Gervasi, B.~Murgante, E.~Stankova, V.~Korkhov, C.~Torre, A.~M.~A. Rocha, D.~Taniar, B.~O. Apduhan, and E.~Tarantino, Eds.\hskip 1em plus 0.5em minus 0.4em\relax Cham: Springer International Publishing, 2019, pp. 613--626.

\bibitem{rateLiteracy}
\BIBentryALTinterwordspacing
W.~P. Review. (2024) Literacy rate by country 2024. [Online]. Available: \url{https://worldpopulationreview.com/country-rankings/literacy-rate-by-country}
\BIBentrySTDinterwordspacing

\bibitem{10.1145/3001913.3001919}
\BIBentryALTinterwordspacing
S.~Castle, F.~Pervaiz, G.~Weld, F.~Roesner, and R.~Anderson, ``Let's talk money: Evaluating the security challenges of mobile money in the developing world,'' in \emph{Proceedings of the 7th Annual Symposium on Computing for Development}, ser. ACM DEV '16.\hskip 1em plus 0.5em minus 0.4em\relax New York, NY, USA: Association for Computing Machinery, 2016. [Online]. Available: \url{https://doi.org/10.1145/3001913.3001919}
\BIBentrySTDinterwordspacing

\bibitem{pousttchi2004assessment}
K.~Pousttchi and M.~Schurig, ``Assessment of today's mobile banking applications from the view of customer requirements,'' in \emph{37th Annual Hawaii International Conference on System Sciences, 2004. Proceedings of the}.\hskip 1em plus 0.5em minus 0.4em\relax IEEE, 2004, pp. 10--pp.

\bibitem{tufano2017and}
M.~Tufano, F.~Palomba, G.~Bavota, R.~Oliveto, M.~Di~Penta, A.~De~Lucia, and D.~Poshyvanyk, ``When and why your code starts to smell bad (and whether the smells go away),'' \emph{IEEE Transactions on Software Engineering}, vol.~43, no.~11, pp. 1063--1088, 2017.

\bibitem{elkhail2019relating}
A.~A. Elkhail and T.~Cerny, ``On relating code smells to security vulnerabilities,'' in \emph{2019 IEEE 5th intl conference on big data security on cloud (BigDataSecurity), IEEE Intl Conference on High Performance and Smart Computing,(HPSC) and IEEE intl conference on intelligent data and security (IDS)}.\hskip 1em plus 0.5em minus 0.4em\relax IEEE, 2019, pp. 7--12.

\bibitem{ghafari2017security}
M.~Ghafari, P.~Gadient, and O.~Nierstrasz, ``Security smells in android,'' in \emph{2017 IEEE 17th international working conference on source code analysis and manipulation (SCAM)}.\hskip 1em plus 0.5em minus 0.4em\relax IEEE, 2017, pp. 121--130.

\bibitem{Bceao}
\BIBentryALTinterwordspacing
BCEAO. (2022, june) Paysage bancaire. [Online]. Available: \url{https://www.bceao.int/fr/content/paysage-bancaire}
\BIBentrySTDinterwordspacing

\bibitem{GS}
\BIBentryALTinterwordspacing
GlobalStats. (2022) Mobile operating system market share in africa. [Online]. Available: \url{https://gs.statcounter.com/os-market-share/mobile/africa/2022}
\BIBentrySTDinterwordspacing

\bibitem{Allix:2016:ACM:2901739.2903508}
\BIBentryALTinterwordspacing
K.~Allix, T.~F. Bissyand{\'e}, J.~Klein, and Y.~Le~Traon, ``Androzoo: Collecting millions of android apps for the research community,'' in \emph{Proceedings of the 13th International Conference on Mining Software Repositories}, ser. MSR '16.\hskip 1em plus 0.5em minus 0.4em\relax New York, NY, USA: ACM, 2016, pp. 468--471. [Online]. Available: \url{http://doi.acm.org/10.1145/2901739.2903508}
\BIBentrySTDinterwordspacing

\bibitem{mYuen}
\BIBentryALTinterwordspacing
M.~Yuen. (2023) Here are the top 50 biggest european banks in 2023. [Online]. Available: \url{https://www.emarketer.com/insights/largest-banks-europe-list/}
\BIBentrySTDinterwordspacing

\bibitem{gaby}
\BIBentryALTinterwordspacing
G.~Villaluz and Z.~Gull. (2023) 50 largest us banks by total assets, q3 2023. [Online]. Available: \url{https://www.spglobal.com/marketintelligence/en/news-insights/latest-news-headlines/50-largest-us-banks-by-total-assets-q3-2023-79625289}
\BIBentrySTDinterwordspacing

\bibitem{bDir}
\BIBentryALTinterwordspacing
B.~Finance. (2023) Banking 500 2023 ranking. [Online]. Available: \url{https://brandirectory.com/rankings/banking/2023/table}
\BIBentrySTDinterwordspacing

\bibitem{10.1145/2259051.2259056}
\BIBentryALTinterwordspacing
A.~Bartel, J.~Klein, Y.~Le~Traon, and M.~Monperrus, ``Dexpler: converting android dalvik bytecode to jimple for static analysis with soot,'' in \emph{Proceedings of the ACM SIGPLAN International Workshop on State of the Art in Java Program Analysis}, ser. SOAP '12.\hskip 1em plus 0.5em minus 0.4em\relax New York, NY, USA: Association for Computing Machinery, 2012, p. 27–38. [Online]. Available: \url{https://doi.org/10.1145/2259051.2259056}
\BIBentrySTDinterwordspacing

\bibitem{lam2011soot}
P.~Lam, E.~Bodden, O.~Lhot{\'a}k, and L.~Hendren, ``The soot framework for java program analysis: a retrospective,'' in \emph{Cetus Users and Compiler Infastructure Workshop (CETUS 2011)}, vol.~15, no.~35, 2011.

\bibitem{10.1145/3540250.3549091}
\BIBentryALTinterwordspacing
S.~Arzt, ``Security code smells in apps: Are we getting better?'' in \emph{Proceedings of the 30th ACM Joint European Software Engineering Conference and Symposium on the Foundations of Software Engineering}, ser. ESEC/FSE 2022.\hskip 1em plus 0.5em minus 0.4em\relax New York, NY, USA: Association for Computing Machinery, 2022, p. 245–255. [Online]. Available: \url{https://doi.org/10.1145/3540250.3549091}
\BIBentrySTDinterwordspacing

\bibitem{vallee1998jimple}
R.~Vallee-Rai and L.~J. Hendren, ``Jimple: Simplifying java bytecode for analyses and transformations,'' 1998.

\bibitem{FlowDroid}
S.~Arzt, S.~Rasthofer, C.~Fritz, E.~Bodden, A.~Bartel, J.~Klein, Y.~Le~Traon, D.~Octeau, and P.~McDaniel, ``Flowdroid: Precise context, flow, field, object-sensitive and lifecycle-aware taint analysis for android apps,'' in \emph{ACM SIGPLAN Notices}, vol.~49, no.~6.\hskip 1em plus 0.5em minus 0.4em\relax ACM, 2014, pp. 259--269.

\bibitem{10.1145/3643991.3644866}
\BIBentryALTinterwordspacing
J.~Samhi, T.~F. Bissyand\'{e}, and J.~Klein, ``Androlibzoo: A reliable dataset of libraries based on software dependency analysis,'' in \emph{Proceedings of the 21st International Conference on Mining Software Repositories}, ser. MSR '24.\hskip 1em plus 0.5em minus 0.4em\relax New York, NY, USA: Association for Computing Machinery, 2024, p. 32â€“36. [Online]. Available: \url{https://doi.org/10.1145/3643991.3644866}
\BIBentrySTDinterwordspacing

\bibitem{owasp1}
\BIBentryALTinterwordspacing
O.~M.~A. Security. Mobile app cryptography. [Online]. Available: \url{https://mas.owasp.org/MASTG/General/0x04g-Testing-Cryptography/#mobile-app-cryptography}
\BIBentrySTDinterwordspacing

\bibitem{guardrails1}
\BIBentryALTinterwordspacing
GUARDRAILS. Insecure algorithm. [Online]. Available: \url{https://docs.guardrails.io/docs/vulnerability-classes/insecure-use-of-crypto/insecure-algorithm}
\BIBentrySTDinterwordspacing

\bibitem{mappings2010cwe}
T.~Mappings, ``Cwe-321: Use of hard-coded cryptographic key,'' \emph{CWE Version 1.11}, vol. 629, p. 420, 2010.

\bibitem{codeQL}
\BIBentryALTinterwordspacing
CodeQL. Android webview settings allows access to content links. [Online]. Available: \url{https://codeql.github.com/codeql-query-help/java/java-android-websettings-allow-content-access/#}
\BIBentrySTDinterwordspacing

\bibitem{svoboda}
\BIBentryALTinterwordspacing
D.~Svoboda. (2014, October) Drd04-j. do not log sensitive information. [Online]. Available: \url{https://wiki.sei.cmu.edu/confluence/display/android/DRD04-J.+Do+not+log+sensitive+information}
\BIBentrySTDinterwordspacing

\bibitem{Bassole}
D.~Bassol{\'e}, G.~Koala, Y.~Traor{\'e}, and O.~Si{\'e}, ``Vulnerability analysis in mobile banking and payment applications on android in african countries,'' in \emph{Innovations and Interdisciplinary Solutions for Underserved Areas}, J.~P.~R. Thorn, A.~Gueye, and A.~P. Hejnowicz, Eds.\hskip 1em plus 0.5em minus 0.4em\relax Cham: Springer International Publishing, 2020, pp. 164--175.

\bibitem{10.1145/3317549.3319723}
\BIBentryALTinterwordspacing
J.~Bowers, I.~N. Sherman, K.~R.~B. Butler, and P.~Traynor, ``Characterizing security and privacy practices in emerging digital credit applications,'' ser. WiSec '19.\hskip 1em plus 0.5em minus 0.4em\relax New York, NY, USA: Association for Computing Machinery, 2019, p. 94–107. [Online]. Available: \url{https://doi.org/10.1145/3317549.3319723}
\BIBentrySTDinterwordspacing

\bibitem{10.1145/3108421.3108433}
\BIBentryALTinterwordspacing
E.-r. Latifa, E.~K.~M. Ahemed, and E.~G. Mohamed, ``Side-effects of permissions requested by mobile banking on android platform: A case study of morocco,'' in \emph{Proceedings of the 1st International Conference on E-Commerce, E-Business and E-Government}, ser. ICEEG '17.\hskip 1em plus 0.5em minus 0.4em\relax New York, NY, USA: Association for Computing Machinery, 2017, p. 76–81. [Online]. Available: \url{https://doi.org/10.1145/3108421.3108433}
\BIBentrySTDinterwordspacing

\bibitem{bojjagani2017vaptai}
S.~Bojjagani and V.~Sastry, ``Vaptai: a threat model for vulnerability assessment and penetration testing of android and ios mobile banking apps,'' in \emph{2017 IEEE 3rd International Conference on Collaboration and Internet Computing (CIC)}.\hskip 1em plus 0.5em minus 0.4em\relax IEEE, 2017, pp. 77--86.

\bibitem{10.1007/978-3-319-28658-7_57}
S.~Bojjagani and V.~N. Sastry, ``Stamba: Security testing for android mobile banking apps,'' in \emph{Advances in Signal Processing and Intelligent Recognition Systems}, S.~M. Thampi, S.~Bandyopadhyay, S.~Krishnan, K.-C. Li, S.~Mosin, and M.~Ma, Eds.\hskip 1em plus 0.5em minus 0.4em\relax Cham: Springer International Publishing, 2016, pp. 671--683.

\bibitem{7947811}
S.~Kaka, V.~N. Sastry, and R.~R. Maiti, ``On the mitm vulnerability in mobile banking applications for android devices,'' in \emph{2016 IEEE International Conference on Advanced Networks and Telecommunications Systems (ANTS)}, 2016, pp. 1--6.

\bibitem{reaves2015mo}
B.~Reaves, N.~Scaife, A.~Bates, P.~Traynor, and K.~R. Butler, ``Mo (bile) money, mo (bile) problems: Analysis of branchless banking applications in the developing world,'' in \emph{24th USENIX Security Symposium (USENIX Security 15)}, 2015, pp. 17--32.

\end{thebibliography}

\end{document}